\documentclass[aps,twocolumn,nofootinbib,groupedaddress,superscriptaddress,longbibliography,notitlepage]{revtex4-2}

\usepackage{amsmath,amssymb}
\usepackage[normalem]{ulem}
\usepackage{graphicx}
\usepackage{longtable,multirow}
\usepackage{mathtools}
\usepackage{dsfont}
\usepackage{amsfonts}
\usepackage{xcolor}
\usepackage{url}
\usepackage{array}
\usepackage{booktabs}

\usepackage[colorlinks=true,linkcolor=teal,citecolor=teal,urlcolor=teal,hypertexnames=false]{hyperref}
\newcommand{\nocontentsline}[3]{}
\newcommand{\tocless}[2]{\bgroup\let\addcontentsline=\nocontentsline#1{#2}\egroup}

\def\ba#1\ea{\begin{align}#1\end{align}}
\def\bg#1\eg{\begin{gather}#1\end{gather}}
\def\bpm{\begin{pmatrix}}
\def\epm{\end{pmatrix}}
\def\bbm{\begin{bmatrix}}
\def\ebm{\end{bmatrix}}

\newcommand{\nn}{\nonumber \\ }
\newcommand{\bb}[1]{{\mathbf #1}}
\newcommand{\bs}[1]{{\boldsymbol #1}}
\newcommand{\bx}{\bb x}
\newcommand{\bk}{\bb k}
\newcommand{\bp}{\bb p}
\newcommand{\bq}{\bb q}

\newcommand{\bR}{\bb R}
\newcommand{\bG}{\bb G}

\newcommand{\td}[1]{\widetilde{#1}}
\newcommand{\cm}{\overline}
\newcommand{\mc}[1]{\mathcal{#1}}

\newcommand{\der}{\partial}
\newcommand{\dg}{\dagger}

\newcommand{\sg}{\sigma}

\newcommand{\vep}{\varepsilon}

\newcommand{\Q}{\mathbb{Q}}

\newcommand{\la}{\leftarrow}

\newcommand{\ket}[1]{|#1\rangle}
\newcommand{\bra}[1]{\langle#1|}
\newcommand{\brk}[2]{\langle#1|#2\rangle}

\newcommand{\gs}{{\rm GS}}

\newcommand{\Ncell}{{N_{\rm cell}}}
\newcommand{\dC}{\Delta \mc{C}_{\rm band}}
\newcommand{\dx}{\Delta x_{\rm band}}

\newcommand{\teal}[1]{{\color{teal} #1}}

\newcommand{\ourtitle}{Stable Wave-Function Zeros Indicate Exciton Topology}

\allowdisplaybreaks

\begin{document}
\title{\ourtitle}

\author{Yoonseok Hwang}
\author{Henry Davenport}
\author{Frank Schindler}
\affiliation{Blackett Laboratory, Imperial College London, London SW7 2AZ, United Kingdom}

\begin{abstract}
Excitons are bound states of electrons and holes whose band topology arises from an interplay between the topology of the underlying electronic bands and the structure of the electron-hole interaction.
In crystalline solids, symmetry representations and topological invariants of the conduction and valence bands constrain the structure of the exciton envelope wave function.
In particular, we show that crystalline symmetry can enforce stable zeros in the exciton wave function. These occur at high-symmetry momenta, including the optically accessible total momentum $p=0$.
We work out how the stable zeros constrain both the relative exciton-band topology (the difference of exciton and non-interacting topological invariants) and the relative band topology (the difference of valence and conduction band invariants), all without requiring detailed knowledge of the band structure or interactions.
We establish these results for two-band excitons in inversion- and rotation-symmetric systems in one and two dimensions, where the relevant topological invariants are the Berry phase in one dimension and the Chern number (modulo the rotation order) in two dimensions.
In two dimensions, the exciton Chern number itself can also be constrained by zero patterns.
\end{abstract}

\maketitle

\let\oldaddcontentsline\addcontentsline
\renewcommand{\addcontentsline}[3]{}

Topology provides a framework for understanding robust properties of electronic states in crystalline solids, such as quantized polarization, bulk topological responses, and protected boundary states~\cite{hasan2010colloquium,qi2011topological,chiu2016classification,shiozaki2016topology,shiozaki2017topological,kruthoff2017topological,shiozaki2022atiyah}.
In band theory, symmetry representations at high-symmetry momenta constrain these topological properties and form the basis of symmetry-based diagnostics of band topology~\cite{fu2007topological,hughes2011inversion,turner2012quantized,alexandradinata2014wilson}.
These approaches, such as topological quantum chemistry~\cite{bradlyn2017topological,cano2018building,elcoro2021magnetic,hwang2026stable} and symmetry indicators~\cite{po2017symmetry,watanabe2018structure,song2018quantitative,khalaf2018symmetry}, make it possible to infer topological properties directly from momentum-space symmetry data, without requiring detailed knowledge of the full wave function throughout the Brillouin zone.
This has enabled a systematic classification of topological phases and facilitated large-scale searches for topological materials, greatly improving the prospects for experimental realization and applications~\cite{zhang2019catalogue,vergniory2019complete,tang2019comprehensive,xu2020high,vergniory2022all}.

An important question is how these ideas extend to interacting quantum states and composite excitations.
Excitons, bound states of electrons and holes, provide a natural setting where topology and interactions coexist~\cite{mott1961transition,jerome1967excitonic}.
The topology of exciton wave functions can arise in different ways: for instance, it may reflect the topology of the underlying electronic bands, or emerge from interaction effects that shape the exciton state.
Recent works have demonstrated that excitons can host nontrivial topology, such as exciton states with finite Chern number, in various systems~\cite{wu2017topological,chen2017chiral,gong2017chiral,blason2020exciton,kwan2021exciton,xie2024theory,xie2024long,froese2025topological,kwan2025textured,cai2025continuum}.
In many cases, such studies start either from topological bands or from engineered interactions that produce topological excitons.
However, a general symmetry-based framework that constrains exciton topology and its relation to the topology of the underlying electronic bands has remained lacking.
In particular, it remains unclear how symmetry alone constrains the structure of exciton wave functions and the topology they encode in a general and model-independent manner.
In this work, we show that crystalline symmetry enforces stable zeros in exciton envelope wave functions. 
The resulting zero patterns diagnose relative topological invariants between the exciton and underlying conduction and valence bands, and, in the two-dimensional (2D) cases considered here, also constrain the individual topological invariants themselves.
Specifically, zero patterns at high-symmetry momenta determine the (i)~\emph{relative exciton-band topology}, i.e., the difference between the topological invariants of the exciton band and the underlying bands, and the (ii)~\emph{relative band topology}, defined as the difference between the topological invariants of the conduction and valence bands.
In 1D inversion-symmetric systems the relevant topological invariants are Berry phases, or equivalently Wannier centers, while in 2D rotation-symmetric systems they are Chern numbers modulo the rotation order.
As a result, both band-induced and interaction-induced exciton topology can be understood and contrasted within a unified framework in terms of symmetry-enforced zeros.
Importantly, the relative band topology can already be inferred from the zero patterns at total exciton momentum $p=0$, which are directly accessible in spectroscopy experiments~\cite{man2021experimental}, providing a practical route to extract band topology through exciton wave functions.
While we focus on excitons in 1D inversion-symmetric and 2D rotation-symmetric systems to establish these results, the underlying symmetry-based mechanism applies more broadly.
Our results establish stable zeros as symmetry-enforced fingerprints of exciton topology and extend symmetry-based diagnostics to the topology of interacting composite excitations.

\teal{\it Symmetry of exciton wave functions---}
We consider excitons formed from a conduction band $c$ and a valence band $v$, whose creation operators at momentum $\bk$ are denoted as $c^\dg_\bk$ and $v^\dg_\bk$.
An exciton state with total momentum $\bp$ (compared to the ground state $\ket{\gs} = \prod_\bk \, v^\dg_\bk \ket{0}$ where $\ket{0}$ is the vacuum state) can be written as
\bg
\ket{\bp_{\rm exc}}
= \sum_\bk \, \phi_\bk (\bp) \, c^\dg_{\bk + \bp} \, v_\bk \, \ket{\gs}
:= \Phi_\bp \ket{\gs},
\label{eq:exc_def}
\eg
where $\phi_\bk (\bp)$ is the exciton envelope wave function (EWF), normalized as $\sum_\bk \, |\phi_\bk (\bp)|^2 = 1$, and periodic in the Brillouin zone (BZ), as discussed in the Supplemental Material (SM)~\cite{supple}.
The operator $\Phi_\bp$ thus creates an exciton with total momentum $\bp$.

Crystalline symmetry relates exciton states at different total momenta.
For a unitary symmetry operation $g$, crystal momentum transforms as $\bp \to g \bp$ modulo a reciprocal lattice vector $\bG$.
For example, inversion $P$ maps $\bp \to -\bp$, while an $n$-fold rotation $C_n$ maps $\bp$ to the rotated momentum.
We define the exciton sewing matrix $\mc B_g (\bp)$ through
\bg
g \, \Phi_\bp \, g^{-1} = \Phi_{g\bp} \, \mc B_g(\bp).
\label{eq:exc_sym}
\eg
Since we consider a single exciton band, $\mc B_g (\bp)$ is a $U(1)$ phase factor.
At high-symmetry momenta (HSMs) satisfying $g \bp_* = \bp_* \pmod{\bG}$, the exciton state transforms into itself, and $\mc B_g (\bp_*)$ reduces to the corresponding symmetry eigenvalue.

Since the exciton state in Eq.~\eqref{eq:exc_def} is built from electron operators in the conduction and valence bands, the symmetry action is determined by the band sewing matrices.
For the conduction band, the sewing matrix $B_{c,g} (\bk)$ is defined by
\bg
g \, c^\dg_\bk \, g^{-1} = c^\dg_{g\bk} \, B_{c,g} (\bk),
\label{eq:band_sewing}
\eg
with an analogous definition for the valence band.
Since each band is nondegenerate, $B_{c/v,g} (\bk)$ are $U(1)$ phase factors.
Further discussion of sewing matrices, including more rigorous notation and antiunitary symmetries, is provided in the SM~\cite{supple}.

Using Eq.~\eqref{eq:band_sewing}, we evaluate the symmetry action on the EWF, obtaining~\cite{supple}
\bg
B_{c,g} (\bk+\bp) B_{v,g} (\bk)^{-1} \, \phi_\bk(\bp)
= \phi_{g\bk} (g\bp) \, \mc B_g (\bp),
\label{eq:exc_sewing_relation}
\eg
which relates the exciton sewing matrix to the band sewing matrices and the EWF.
In particular, for $\bk_*$ and $\bp_*$ that are HSMs of $g$, the sewing matrices reduce to symmetry eigenvalues, and Eq.~\eqref{eq:exc_sewing_relation} constrains the EWF through symmetry eigenvalues of the exciton state and the underlying bands.
A symmetry relation equivalent to Eq.~\eqref{eq:exc_sewing_relation} was also obtained in Ref.~\cite{nalabothula2025symmetries}.
However, the consequences of this relation for the structure and topology of EWFs have not been explored.
As we show below, Eq.~\eqref{eq:exc_sewing_relation} can enforce stable zeros in EWFs that diagnose exciton topology. 

\teal{\it Stable zeros and exciton topology---}
We now show that Eq.~\eqref{eq:exc_sewing_relation} can enforce stable zeros in EWFs.
These zeros are determined entirely by symmetry eigenvalues and therefore represent robust features of the EWF.
We illustrate this mechanism explicitly using a gapped exciton in a 1D inversion-symmetric system with inversion symmetry $P$.
The $P$ sewing matrices for the conduction band, valence band, and exciton at $P$-invariant HSMs $k_*, p_* \in \{0,\pi\}$ reduce to inversion eigenvalues,
\bg
B_{c/v, P} (k_*) = \xi_{c/v} (k_*),
\quad
\mc B_P (p_*) = \xi_{\rm exc} (p_*),
\eg
where $\xi_{c/v/{\rm exc}} \in \{+1, -1\}$.
Applying Eq.~\eqref{eq:exc_sewing_relation} at HSMs gives
\bg
\xi_c (k_*+p_*) \, \xi_v (k_*)^{-1} \, \phi_{k_*} (p_*)
= \phi_{k_*} (p_*) \, \xi_{\rm exc} (p_*).
\label{eq:inv_constraint}
\eg

Crucially, if $\xi_c (k_*+p_*) \, \xi_v (k_*)^{-1} \neq \xi_{\rm exc} (p_*)$, the only solution to Eq.~\eqref{eq:inv_constraint} is $\phi_{k_*} (p_*) = 0$.
This zero is enforced by symmetry and cannot be removed without breaking $P$ or closing the exciton gap.
It is gauge invariant, since gauge transformations multiply the wave function only by a phase~\cite{supple}.
It therefore represents a stable, symmetry-protected feature of the EWF.
Since both the zero condition and exciton topology in this system are determined by $P$ eigenvalues, the resulting EWF stable zeros encode the relative topology between exciton and underlying bands.

This connection can be understood from the Wannier representation~\cite{marzari2012maximally}.
For 1D $P$-symmetric systems, inversion eigenvalues determine and quantize the Wannier centers of electronic bands~\cite{fu2007topological,hughes2011inversion,turner2012quantized,alexandradinata2014wilson} and those of excitons~\cite{haber2023maximally,davenport2024interaction,davenport2026exciton, davenport2026berry}.
Explicitly,
\bg
(-1)^{2 x_n} = \xi_n (0) \,\xi_n (\pi)^{-1}
\quad (n = c, v, {\rm exc}),
\label{eq:1d_center}
\eg
where $x_n \in \{0, 1/2\}$ is the Wannier center in lattice units.
Physically, the exciton Wannier center $x_{\rm exc}$ is equivalently given by the exciton Berry phase divided by $2\pi \pmod 1$~\cite{davenport2024interaction,davenport2026exciton, davenport2026berry}.
This Berry phase can be computed from exciton Berry connections constructed from either the projected electron or hole position operators, and the two coincide in the presence of $P$ symmetry~\cite{davenport2024interaction,davenport2026exciton, davenport2026berry}.
%

\begin{figure}[t]
\centering
\includegraphics[width=0.45\textwidth]{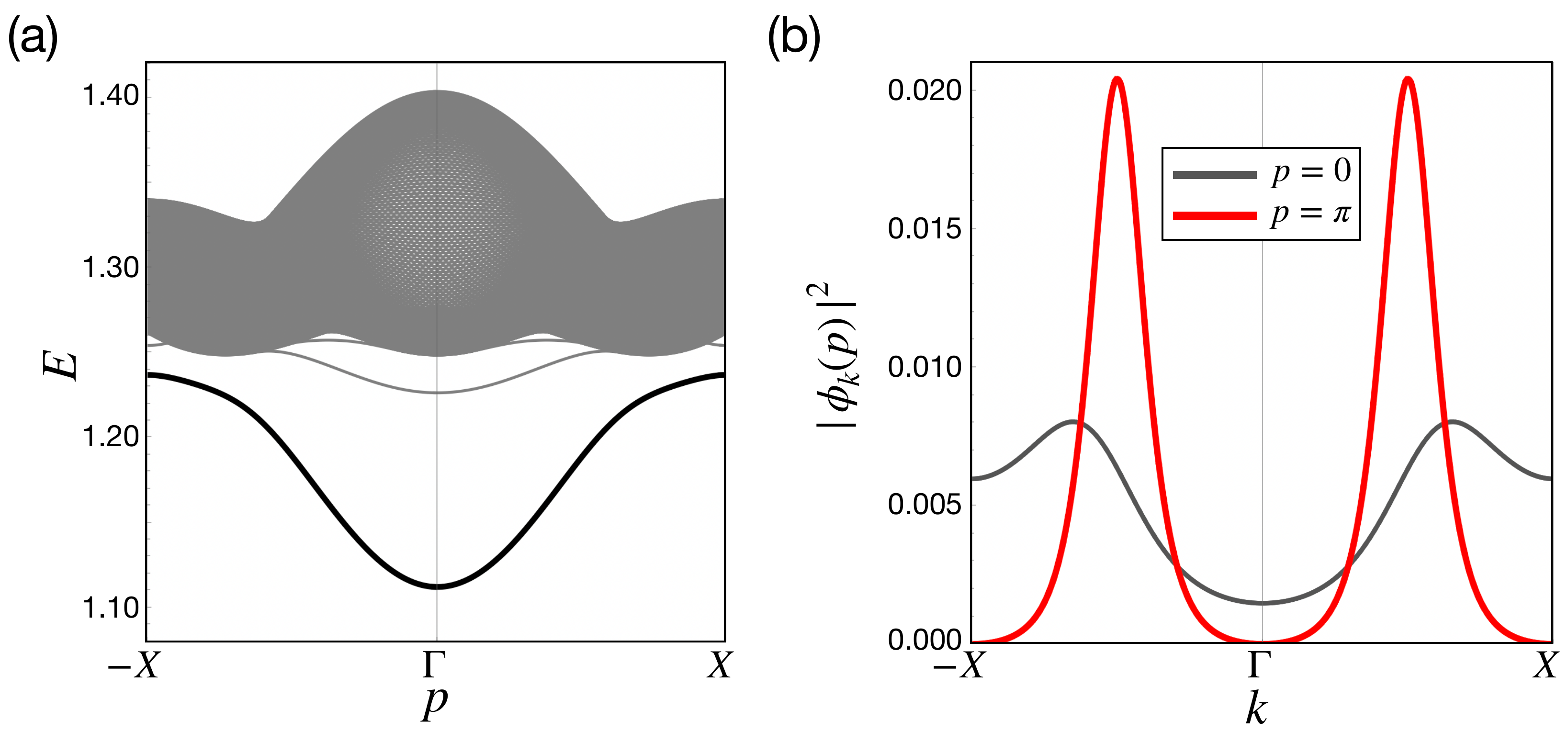}
\caption{Excitons in a one-dimensional inversion-symmetric lattice model.
(a) Exciton spectrum as a function of total momentum $p$ (see SM~\cite{supple} for model details).
The lowest gapped exciton band is shown in black.
In this model, the lowest exciton band has Wannier center $x_{\rm exc}=1/2$, while the conduction and valence bands have $x_{c/v}=0$.
(b) EWF intensity $|\phi_k(p)|^2$ of the lowest exciton in (a), shown for $p=0$ (dark gray) and $p=\pi$ (red).
The stable zero pattern agrees with $s_c=s_v=1/2$ (see Tab.~\ref{table:1d_pattern}).}
\label{fig:1d_model}
\end{figure}

To relate stable zeros to Wannier centers, we evaluate Eq.~\eqref{eq:inv_constraint} at $p_*=0$ and $\pi$:
\ba
\xi_c (k_*) \, \phi_{k_*} (0)
=& \xi_v (k_*) \, \xi_{\rm exc} (0) \,\phi_{k_*} (0),
\label{eq:inv_constraint2}
\\
\xi_c (k_*+\pi) \,\phi_{k_*} (\pi)
=& \xi_v (k_*) \,\xi_{\rm exc} (\pi) \, \phi_{k_*} (\pi).
\label{eq:inv_constraint3}
\ea
From these equations, we show in the following that the pattern of symmetry-enforced zeros uniquely determines the Wannier center shifts between the exciton and the bands,
$s_{c/v} = x_{c/v} - x_{\rm exc}\pmod{1}$.

From Eqs.~\eqref{eq:inv_constraint2} and \eqref{eq:inv_constraint3}, $\phi_{k_*} (0)$ is a stable zero if and only if $\xi_c (k_*) = -\xi_v (k_*) \xi_{\rm exc}(0)$, and similarly $\phi_{k_*} (\pi)$ is a stable zero if and only if $\xi_c (k_*+\pi) = -\xi_v (k_*) \xi_{\rm exc}(\pi)$.
Thus, when both $\phi_{k_*}(0)$ and $\phi_{k_*}(\pi)$ are stable zeros, these conditions imply, upon taking their ratio,
\bg
\xi_c(k_*) \xi_c (k_* +\pi)^{-1} = \xi_{\rm exc} (0) \xi_{\rm exc} (\pi)^{-1}.
\label{eq:inv_sc_example}
\eg
Hence, from Eq.~\eqref{eq:1d_center}, $x_c = x_{\rm exc} \pmod 1$ so that $s_c = 0 \pmod 1$.
On the other hand, if both $\phi_{k_*} (0)$ and $\phi_{k_*} (\pi)$ are nonzero,
the constraints take the form $\xi_c (k_*) = \xi_v (k_*) \xi_{\rm exc}(0)$ and $\xi_c (k_* + \pi) = \xi_v (k_*) \xi_{\rm exc}(\pi)$, which again imply Eq.~\eqref{eq:inv_sc_example} and thus $s_c=0$.
Therefore, when $s_c=0$, $\phi_{k_*}(0)$ and $\phi_{k_*}(\pi)$ must either both vanish or both be nonzero at $k_*=0$ and $\pi$, as shown in Tab.~\ref{table:1d_pattern}.
Here and below, ``nonzero" means not required to vanish by symmetry, although accidental zeros may occur.

If instead exactly one of $\phi_{k_*}(0)$ or $\phi_{k_*}(\pi)$ is a stable zero, then $\xi_c (k_*) \xi_c (k_*+\pi)^{-1} = - \xi_{\rm exc} (0) \xi_{\rm exc} (\pi)^{-1}$, which from Eq.~\eqref{eq:1d_center} implies $x_c = x_{\rm exc} + 1/2 \pmod 1$ and thus $s_c=1/2 \pmod 1$.
A completely analogous argument applies to the Wannier-center shift between valence band and exciton, $s_v = x_v - x_{\rm exc} \pmod 1$.
If $s_v = 0$, the symmetry constraints are identical, and $\phi_{k_*+\pi} (0)$ and $\phi_{k_*} (\pi)$ must either both be nonzero or both vanish.
If instead $s_v = 1/2$, the symmetry constraints differ by a minus sign, and exactly one of $\phi_{k_*+\pi} (0)$ and $\phi_{k_*} (\pi)$ must vanish.
Further details of the derivation are provided in the SM~\cite{supple}.

As an important example, consider $(s_c, s_v)=(1/2, 1/2)$, corresponding to interaction-induced exciton topology relative to the underlying bands.
In this case, there are two possible stable-zero patterns for $\phi_k (p)$ at $P$-invariant HSMs,
\ba
\bbm
\phi_0 (0) & \phi_\pi (0) \\
\phi_0 (\pi) & \phi_\pi (\pi)
\ebm
= \bbm \bullet & \bullet \\ 0 & 0 \ebm
\, \text{or} \,
\bbm 0 & 0 \\ \bullet & \bullet \ebm,
\label{eq:zero_def}
\ea
up to accidental zeros, where $\bullet$ denotes entries that are generically nonzero.

More generally, in 1D $P$-symmetric systems, a given stable-zero pattern \emph{uniquely} determines the pair $(s_c,s_v)$.
Since the relative band topology is defined by the band Wannier-center difference $\dx = x_c - x_v = s_c - s_v \pmod 1$, stable zeros encode both the relative exciton-band topology and the relative band topology.
The complete correspondence between zero patterns and $(s_c,s_v)$ is summarized in Tab.~\ref{table:1d_pattern}.

We verify this prediction in a lattice model with inversion symmetry, shown in Fig.~\ref{fig:1d_model}.
The model and computational details are provided in the SM~\cite{supple}.
The valence and conduction bands satisfy $\xi_v (0) = \xi_v (\pi) = -1$ and $\xi_c (0) = \xi_c (\pi) = +1$.
For the lowest gapped exciton shown in Fig.~\ref{fig:1d_model}(a), the exciton inversion eigenvalues are $\xi_{\rm exc} (0) = -1$ and $\xi_{\rm exc} (\pi) = +1$, which corresponds to stable zeros only at $p = \pi$, i.e., $\phi_0 (\pi) = \phi_\pi (\pi) = 0$.
Conversely, the observed zero pattern uniquely determines
$(s_c,s_v)=(1/2,1/2)$ and $\dx=0$, consistent with the relation derived above.

\teal{\it Zeros at $p=0$ as band topology probes---}
The results above show that stable-zero patterns determine both the relative exciton-band topology $(s_c,s_v)$ and the relative band topology $\dx$.
Remarkably, even partial information about the zero pattern at $p=0$ suffices to constrain $\dx$.
In optical spectroscopy, excitons at total momentum $p=0$ are directly accessible~\cite{man2021experimental}.
This motivates examining how much topological information can be extracted from the zero structure at $p=0$ alone.

In 1D inversion-symmetric systems, the zero pattern at $p=0$ is determined by the pair of amplitudes $\phi_0(0)$ and $\phi_\pi(0)$.
Even without knowledge of the $p=\pi$ sector, these quantities already constrain $\dx$, which quantifies the relative band topology between the conduction and valence bands.
From Tab.~\ref{table:1d_pattern}, we conclude that:
(i) If both $\phi_0(0)$ and $\phi_\pi(0)$ are nonzero, or if both vanish, then $\dx = 0 \pmod 1$.
(ii) If exactly one of $\phi_0(0)$ or $\phi_\pi(0)$ vanishes, then $\dx = 1/2 \pmod 1$, indicating nontrivial relative band topology.

Thus, stable exciton zeros at $p=0$ provide an experimentally accessible diagnosis of the relative topology of the underlying non-interacting bands.
In particular, the mere existence of a gapped exciton, together with knowledge of its $p=0$ zero pattern, suffices to diagnose whether the underlying bands have nontrivial relative topology.
This observation provides a method to access band topology through optical responses.

\newcommand{\mcell}[1]{\makebox[1.0em][c]{$#1\vphantom{0}$}}
\newcommand{\patmat}[4]{%
\bbm
\mcell{#1} & \mcell{#2} \\
\mcell{#3} & \mcell{#4}
\ebm
}

{\renewcommand{\arraystretch}{1.4}
\begin{table}[t!]
\centering
\caption{Correspondence of stable-zero patterns of $\phi_k(p)$ in Eq.~\eqref{eq:zero_def} and Wannier-center shifts $(s_c,s_v)$ in 1D inversion-symmetric systems.
A given zero pattern determines $(s_c,s_v)$ and the band Wannier-center difference $\dx = x_c - x_v = s_c - s_v \pmod 1$.
Here $0$ denotes symmetry-enforced zeros, while the symbol $\bullet$ indicates entries that are generically nonzero. Each entry of the $2\times 2$ pattern corresponds to the EWF amplitude $\phi_{k_*}(p_*)$ at an inversion-invariant HSM, with rows labeled by $p_*=0,\pi$ (top to bottom) and columns labeled by $k_*=0,\pi$ (left to right), matching Eq.~\eqref{eq:zero_def}.}
\label{table:1d_pattern}
\begin{tabular}{c|c|c@{, \hspace{8pt}}c}
\hline \hline
$(s_c,s_v)$ & $\dx$ & \multicolumn{2}{c}{Stable-zero pattern} \\
\hline
$(0,0)$ & $0$ &
$\patmat{\bullet}{\bullet}{\bullet}{\bullet}$ &
$\patmat{0}{0}{0}{0}$
\\
\noalign{\hrule height 0.1pt}
$(\tfrac 12,\tfrac 12)$ & $0$ &
$\patmat{\bullet}{\bullet}{0}{0}$&
$\patmat{0}{0}{\bullet}{\bullet}$
\\
\noalign{\hrule height 0.1pt}
$(\tfrac 12,0)$ & $\tfrac 12$ &
$\patmat{\bullet}{0}{0}{\bullet}$ &
$\patmat{0}{\bullet}{\bullet}{0}$
\\
\noalign{\hrule height 0.1pt}
$(0,\tfrac 12)$ & $\tfrac 12$ &
$\patmat{\bullet}{0}{\bullet}{0}$ &
$\patmat{0}{\bullet}{0}{\bullet}$
\\
\hline\hline
\end{tabular}
\end{table}}

\teal{\it Rotation-symmetric excitons in 2D---}
The analysis above for 1D inversion symmetry can be generalized to higher dimensions.
The symmetry constraint in Eq.~\eqref{eq:exc_sewing_relation} is completely universal and applies to crystalline symmetries in arbitrary dimensions.
We now show that the same zero-based logic extends naturally to two-dimensional systems with $C_n$ rotation ($n=2,3,4,6$) in the absence of time-reversal symmetry.

For electronic bands with $C_n$ symmetry, it is well known that the Chern number $\mc C$ modulo $n$ is determined by rotation eigenvalues at the HSMs~\cite{fang2012bulk}.
A key technical ingredient of our work is the symmetry transformation of exciton Wilson loops.
Using the exciton Wilson-loop formulation introduced in Ref.~\cite{davenport2026exciton}, we derive their transformation under crystalline symmetries in terms of the exciton sewing matrix, in a form directly analogous to the electronic case.
As one consequence, the exciton Chern number~\cite{davenport2026exciton} is determined modulo $n$ by rotation eigenvalues at HSMs.
Specifically,
\bg
\prod_{\bp_* \in {\rm HSM}} \mc B_{g_{\bp_*}} (\bp_*)
= e^{\frac{2 \pi i}{n} \mc C_{\rm exc}},
\label{eq:exc_chern}
\eg
where $g_{\bp_*}$ denotes the highest-order rotation symmetry that leaves $\bp_*$ invariant.
For example, in a $C_4$-symmetric system one has 
$\mc B_{C_4}(\Gamma) \,\mc B_{C_4}(M) \,\mc B_{C_2}(X)
= e^{i \frac{\pi}{2} \mc C_{\rm exc}}$.
Thus $\mc C_{\rm exc} \pmod n$ is fixed by rotation eigenvalues at the HSMs, as in the electronic case.
Our HSM notation follows standard conventions; for example, $M=(\pi,\pi)$ and $X=(\pi,0)$ for $C_4$.
The derivation of Eq.~\eqref{eq:exc_chern} and the corresponding Wilson-line identities are presented in the SM~\cite{supple}.

Denoting the Chern numbers of the conduction and valence bands by $\mc C_c$ and $\mc C_v$, the relative band topology is characterized by the relative band Chern number $\dC = \mc C_c - \mc C_v \pmod n$, and we define the Chern-number shifts $\mc S_{c/v} = \mc C_{c/v} - \mc C_{\rm exc} \pmod n$ as indicators of relative exciton-band topology.
Equation~\eqref{eq:exc_sewing_relation} imposes symmetry constraints relating the full zero pattern of the EWF $\phi_\bk (\bp)$ at all rotation-invariant HSMs $(\bk_*, \bp_*)$ to the shifts $\mc S_{c/v}$ between the exciton band and the underlying bands, and even to the Chern numbers $\mc C_{{\rm exc}/c/v}$ themselves modulo $n$.
For $C_2$, a given zero pattern uniquely determines $\mc S_{c/v}$ modulo $2$, and hence $\dC$, in direct analogy to the 1D $P$-symmetric case.
In fact, it also uniquely determines the Chern numbers $\mc C_{{\rm exc}/c/v} \pmod 2$, so that explicit counting rules can be formulated in terms of the zero pattern.
For example, $\mc C_{\rm exc} = N_{(\Gamma, \Gamma)} + N_{(\Gamma, X)} + N_{(Y, Y)} + N_{(Y, M)} \pmod 2$, where $N_{(\bk_*, \bp_*)}=1$ if $\phi_{\bk_*} (\bp_*)$ is a stable zero, and $0$ otherwise~\cite{supple}.
For $C_{3,4,6}$, zero patterns do not in general uniquely determine the topological invariants, and are less restrictive for the pair $(\mc S_c, \mc S_v)$, but still impose strong constraints on $\dC$ and the individual Chern numbers modulo $n$.
A detailed discussion on zero patterns and explicit examples is provided in the SM~\cite{supple}.

In practice, stable zeros at $\bp=\Gamma=(0,0)$ can be used as a proxy, since the number of possible patterns is smaller and their interpretation is more direct.
These $p=0$ zero patterns already constrain the relative band Chern number $\dC$ modulo $n$.
Table~\ref{table:2d_pattern} summarizes the correspondence between $p=0$ zero patterns and $\dC$ for $C_n$.
We list only those patterns that enforce $\dC \ne 0 \pmod n$.
The complete classification is provided in the SM~\cite{supple}.
Taken together, these results show that EWF zeros provide a practically accessible route to extracting symmetry-constrained topological information about both the exciton and the underlying electronic structure, establishing a unified symmetry-based diagnostic framework for excitons in crystalline systems.

{\renewcommand{\arraystretch}{1.3}
\begin{table}[t!]
\centering
\caption{Zero patterns of the EWF at $\bp=\bb 0$ and relative band Chern number $\dC \pmod n$ for $C_n$.
The symbol $\bullet$ denotes generically nonzero entries.
Only patterns enforcing nontrivial $\dC$ are shown.
The condition $\dC \ne 0$ indicates that $\dC$ may take any nonzero value modulo $n$.
We define $\bs \phi (\bb 0)$ as $[\phi_\Gamma (\Gamma), \phi_X (\Gamma), \phi_Y (\Gamma), \phi_M (\Gamma)]$ for $C_2$; $[\phi_\Gamma (\Gamma), \phi_K (\Gamma), \phi_{K'} (\Gamma)]$ for $C_3$; $[\phi_\Gamma (\Gamma), \phi_M (\Gamma), \phi_X (\Gamma)]$ for $C_4$; and $[\phi_\Gamma (\Gamma), \phi_K (\Gamma), \phi_M (\Gamma)]$ for $C_6$.
HSM labels follow standard convention.
Precise definitions are given in the SM~\cite{supple}.}
\label{table:2d_pattern}
\begin{tabular}{c|c|c}
\hline\hline
Symmetry& Zero pattern of $\bs \phi (\bb 0)$ & Constraint on $\dC$ \\
\hline
$C_2$ & odd number of zeros & $\dC=1$ \\
\hline
$C_3$ & single zero & $\dC \ne 0$ \\
\hline
\multirow{3}*{$C_4$} & $(\bullet,\bullet,0) $ & $\dC=2$ \\
& $(\bullet,0,\bullet)$ & $\dC \ne 0$ \\
& $(0,\bullet,\bullet)$ & $\dC \ne 0$ \\
\hline
\multirow{4}*{$C_6$} & $(\bullet,\bullet,0)$ & $\dC = 3$ \\
& $(0,\bullet,\bullet)$ & $\dC \ne 0$ \\
& $(\bullet,0,0)$ & $\dC \in \{1,5\}$ \\
& $(\bullet,0,\bullet)$ & $\dC \in \{2,4\}$ \\
\hline\hline
\end{tabular}
\end{table}}

\teal{\it Discussion---}
We have shown that stable, symmetry-enforced zeros of EWFs impose strong constraints on both the relative exciton-band topology and the relative band topology.
In 1D with $P$ symmetry, the full zero pattern uniquely determines the exciton-to-band Wannier-center shift and the band Wannier-center difference, concomitant with the associated Berry phases.
In 2D $C_n$-symmetric systems with broken time-reversal symmetry, stable zeros impose strong constraints on the exciton-to-band Chern-number shifts, the relative band Chern number, and the individual Chern numbers (modulo $n$).
These cases exhaust all 1D and 2D space groups, namely the 1D line group $p1m1$ and the 2D wallpaper group $pn$, for which the relevant invariants are constrained by symmetry eigenvalues~\cite{po2017symmetry}.
A central feature of our approach is that these constraints follow purely from symmetry, independent of microscopic interaction details.
Our results suggest the possibility of broader frameworks for excitons and other interacting composite excitations, analogous to topological quantum chemistry~\cite{bradlyn2017topological,cano2018building,elcoro2021magnetic,hwang2026stable} and symmetry indicators~\cite{po2017symmetry,watanabe2018structure,song2018quantitative,khalaf2018symmetry}.

From an experimental perspective, excitons at total momentum $p=0$ are directly accessible in spectroscopy~\cite{man2021experimental}.
Our results show that stable zero patterns at $p=0$ provide a practically accessible route to extracting symmetry-constrained information about exciton and band topology~\cite{man2021experimental}.
As experimental probes of EWFs continue to advance, combining symmetry analysis with spectroscopic measurements may enable direct identification of interaction-induced topological shifts and relative band topology.

Although we have focused on gapped two-band excitons and unitary crystalline symmetries, our logic and formalism apply more broadly.
For instance, in $C_n$-symmetric systems with spinless time-reversal symmetry, symmetry indicators are known to diagnose nodal structures in electronic systems; exciton zeros may play a similar role in identifying nodal structures in exciton bands~\cite{kim2015dirac,ono2018unified,song2018diagnosis}.
More generally, extending this framework to multiband settings, exploring its relation to exciton quantum geometry~\cite{jankowski2025excitonic,thompson2025topologically}, and applying it to other composite excitations such as trions remain important directions for future work.

\begin{acknowledgments}
This work was supported by a UKRI Future Leaders Fellowship MR/Y017331/1. HD acknowledges support from the Engineering and Physical Sciences Research Council (grant number EP/W524323/1).
\end{acknowledgments}

\bibliography{refs.bib}

\let\addcontentsline\oldaddcontentsline
\clearpage

\onecolumngrid
\begin{center}
\textbf{\large Supplemental Material: \\
\ourtitle}
\end{center}

\setcounter{section}{0}
\setcounter{figure}{0}
\setcounter{equation}{0}
\renewcommand{\thefigure}{S\arabic{figure}}
\renewcommand{\theequation}{S\arabic{equation}}
\renewcommand{\thesection}{S\arabic{section}}

\tableofcontents
\hfill \\

\section{Notation and symmetry conventions for electronic bands}
\label{app:review}

\subsection{Tight-binding notation}
\label{app:notation}
We consider a non-interacting tight-binding model with a finite-dimensional basis.
In real space, the basis is spanned by (L\"owdin) basis orbitals whose creation and annihilation operators are denoted by $c^\dg_{\bR,i}$ and $c_{\bR,i}$, where $\bR$ labels the unit cell and $i=1,\dots,N_{\rm tot}$ labels the orbital within the unit cell.
The orbital corresponding to $c^\dg_{\bR,i}$ is localized at $\bR + \bx_i$, where $\bx_i$ denotes the intracell position.
We define the Fourier transform as
\bg
c^\dg_{\bk,i} = \frac{1}{\sqrt \Ncell} \sum_\bR \, e^{i \bk \cdot (\bR + \bx_i)} c^\dg_{\bR,i}, 
\quad
c_{\bk,i} = \frac{1}{\sqrt \Ncell} \sum_\bR \, e^{-i \bk \cdot (\bR + \bx_i)} c_{\bR,i},
\eg
where $\Ncell$ is the number of unit cells under periodic boundary conditions.

The Hamiltonian and band eigenstates are given by
\bg
\hat H = \sum_{\bk,i,j} \, H(\bk)_{ij} c^\dg_{\bk,i} c_{\bk,j},
\quad
H(\bk) \ket{u_{\bk,n}} = \vep_{\bk,n} \ket{u_{\bk,n}},
\label{seq:tbH}
\eg
where $n$ labels the bands and $\vep_{\bk,n}$ is the band dispersion of the $n$th band.
In components, $\sum_{j=1}^{N_{\rm tot}} \, H(\bk)_{ij} (u_{\bk,n})_j = \vep_{\bk,n} (u_{\bk,n})_i$ with $i,j = 1,\dots, N_{\rm tot}$.
The eigenvectors satisfy the orthonormality condition
\bg
\brk{u_{\bk,n}}{u_{\bk,m}} = \sum_{i=1}^{N_{\rm tot}} \, (u_{\bk,n})_i^* (u_{\bk,m})_i = \delta_{nm}.
\eg
The corresponding eigenstates of $\hat H$ are the Bloch states defined as
\bg
\hat H \ket{\psi_{\bk,n}} = \vep_{\bk,n} \ket{\psi_{\bk,n}}, 
\quad
\ket{\psi_{\bk,n}} = \sum_{i=1}^{N_{\rm tot}} \, (u_{\bk,n})_i \, c^\dg_{\bk,i} \ket{0},
\eg
where $\ket{0}$ is the fermionic vacuum.
This defines the band-basis operators
\bg
c^\dg_{\bk,n} = \sum_{i=1}^{N_{\rm tot}} (u_{\bk,n})_i \, c^\dg_{\bk,i}
= \frac{1}{\sqrt \Ncell} \sum_{\bR,i} \, (u_{\bk,n})_i \, e^{i \bk \cdot (\bR + \bx_i)} \, c^\dg_{\bR,i}.
\eg
Conversely,
\bg
c^\dg_{\bk,i} = \sum_n \, (u_{\bk,n})_i^* \, c^\dg_{\bk,n}, 
\quad
c^\dg_{\bR,i} = \frac{1}{\sqrt{\Ncell}} \sum_{\bk,n} \, (u_{\bk,n})_i^* \, e^{-i \bk \cdot (\bR + \bx_i)} \, c^\dg_{\bk,n}.
\label{seq:band_cr_def}
\eg

We work in the periodic gauge~\cite{alexandradinata2014spin,vanderbilt2018berry}.
Since $c^\dg_{\bk + \bG,i} = e^{i \bG \cdot \bx_i} \, c^\dg_{\bk,i}$, the Hamiltonian $H(\bk)$ transforms as
\bg
H(\bk + \bG)_{ij} = e^{-i \bG \cdot \bx_i} \, H(\bk)_{ij} \, e^{i \bG \cdot \bx_j},
\eg
and hence the eigenstates are not strictly periodic when the intracell positions $\bx_i$ are nontrivial~\cite{shiozaki2017topological}.
We therefore impose the periodic gauge condition
\bg
\ket{u_{\bk + \bG,n}}_i = e^{-i \bG \cdot \bx_i} \ket{u_{\bk,n}}_i,
\label{seq:peri_gauge_1}
\eg
or equivalently $\ket{u_{\bk + \bG,n}} = V(-\bG) \ket{u_{\bk,n}}$ with $V(\bk)_{ij}=e^{i \bk \cdot \bx_i} \delta_{ij}$.
With this choice, the Bloch states and band operators are periodic:
\bg
\ket{\psi_{\bk + \bG,n}} = \ket{\psi_{\bk,n}}, 
\quad
c^\dg_{\bk + \bG,n} = c^\dg_{\bk,n}, 
\quad
c_{\bk + \bG,n} = c_{\bk,n}.
\label{seq:peri_gauge_2}
\eg
While the periodic gauge need not be imposed, it provides a framework in which gauge-invariant quantities, such as symmetry eigenvalues of bands, can be computed unambiguously, as discussed in Sec.~\ref{app:band_sewing}.

\subsection{Symmetry and Sewing matrices for electronic bands}
\label{app:band_sewing}
We now define symmetry operations and sewing matrices, following standard conventions.
Detailed treatments can be found, for example, in Refs.~\cite{alexandradinata2016topological,shiozaki2017topological,alexandradinata2018no,hwang2026building}.
Consider a space-group operation $g=\{O_g|\bb v_g\}$, acting as $\bR \to O_g \bR + \bb v_g$ and $\bk \to g \bk = (-1)^{\tau_g} O_g \bk$, where $\tau_g=0$ (1) for unitary (antiunitary) symmetries.
Here $O_g \in O(d)$ is an orthogonal transformation in $d$ dimensions, and $\bb v_g$ is the (possibly fractional) translation, which can be expressed in the primitive lattice basis with rational coefficients in $\Q^d$.
For example, inversion corresponds to $O_g = -\mathds{1}_d$ (the $d\times d$ identity matrix), $\bb v_g=0$, and $\tau_g=0$, giving $\bk \to -\bk$, while time-reversal symmetry has $O_g=\mathds{1}_d$ and $\tau_g=1$, which also maps $\bk \to -\bk$.

The action of $\hat g$ on the creation operators is
\bg
\hat{g} \, c^\dg_{\bR,i} \, \hat{g}^{-1} = \sum_j \, c^\dg_{\bR',j} \, [U_g]_{ji},
\quad
\hat{g} \, c_{\bR,i} \, \hat{g}^{-1} = \sum_j \, [U_g^{-1}]_{ij} \, c_{\bR',j},
\label{seq:sym_cr_op_1}
\\
\hat{g} \, c^\dg_{\bk,i} \, \hat{g}^{-1} = \sum_j \, c^\dg_{g\bk,j} \, [U_g]_{ji} \, e^{-i g\bk \cdot \bb v_g},
\quad
\hat{g} \, c_{\bk,i} \, \hat{g}^{-1} = \sum_j \, e^{i g\bk \cdot \bb v_g} \, [U_g^{-1}]_{ij} \, c_{g\bk,j},
\label{seq:sym_cr_op_2}
\eg
where $\bR'$ and $j$ are defined by $\bR' + \bx_j = g \circ (\bR + \bx_i) = O_g \bR + O_g \bx_i + \bb v_g$.
When $g$ is antiunitary, we have
$\hat g x \hat g^{-1} = \mc{K} x \mc{K} = x^*$
for any complex-valued quantity $x$ (including numbers, vectors, and matrices), where $\mc{K}$ denotes complex conjugation.
In particular, $\hat g i \hat g^{-1} = (-1)^{\tau_g} i$.
In the band basis,
\bg
\hat{g} \, c^\dg_{\bk,n} \, \hat{g}^{-1} = \sum_m \, c^\dg_{g\bk,m} \, [B_g(\bk)]_{mn},
\quad
\hat{g} \, c_{\bk,n} \, \hat{g}^{-1} = \sum_m \, [B_g(\bk)^{-1}]_{nm} \, c_{g\bk,m},
\label{seq:sym_cr_op_3}
\eg
which defines the sewing matrix $B_g(\bk)$, which is unitary.

Combining Eqs.~\eqref{seq:band_cr_def}, \eqref{seq:sym_cr_op_2} and \eqref{seq:sym_cr_op_3}, we obtain
\bg
\sum_j \, e^{- i g \bk \cdot \bb v_g} \, [U_g]_{ij} \cm{(u_{\bk,n})_j}^{\tau_g} = (u_{g \bk,m})_i \, [B_g(\bk)]_{mn},
\label{seq:sewing_u}
\eg
where we introduced the notation
\bg
\cm{x}^{\tau_g} := \mc K^{\tau_g} \, x \, \mc K^{\tau_g}= \begin{cases}
x & \text{ for } \tau_g=0, \\
x^* & \text{ for } \tau_g=1.
\end{cases}
\eg
Note that $\mc{K}^1=\mc{K}$ and $\mc{K}^0$ is the identity.
Using the orthonormality of $\ket{u_{\bk,n}}$, we find
\bg
[B_g(\bk)]_{nm} = \bra{u_{g \bk,n}} e^{- i g \bk \cdot \bb v_g} \, U_g \mc{K}^{\tau_g} \ket{u_{\bk,m}} \, \mc{K}^{\tau_g}.
\eg
For unitary symmetries ($\tau_g=0$), this reduces to
\bg
[B_g(\bk)]_{nm}
= e^{-i g\bk \cdot \bb v_g} \bra{u_{g\bk,n}} U_g \ket{u_{\bk,m}}.
\label{seq:sewing_uni}
\eg
For antiunitary symmetries ($\tau_g=1$), we obtain
\bg
[B_g(\bk)]_{nm}
= e^{-i g\bk \cdot \bb v_g} \bra{u_{g\bk,n}} U_g \cm{\ket{u_{\bk,m}}}.
\eg
In the periodic gauge, $B_g(\bk)$ is periodic in the Brillouin zone.
At high-symmetry momenta $\bk_*$ satisfying $g \bk_*=\bk_* \pmod{\bG}$, where $\bG$ is a reciprocal lattice vector, and for unitary symmetries, the eigenvalues of $B_g (\bk_*)$ give the symmetry eigenvalues of the bands, and its trace equals the group-theoretical character.
Under the periodic gauge in Eq.~\eqref{seq:peri_gauge_1}, Eq.~\eqref{seq:sewing_uni} reduces to
\bg
[B_g(\bk_*)]_{nm}
= e^{-i g \bk_* \cdot \bb v_g}
\bra{u_{\bk_*,n}} V(g\bk_* - \bk_*) \, U_g \ket{u_{\bk_*,m}}.
\eg
In this form, $B_g(\bk_*)$ is computed entirely from the eigenstates at $\bk_*$.
Under a gauge transformation $\ket{u_{\bk,n}} \to \sum_m \, \ket{u_{\bk,m}} \, G(\bk)_{mn}$ with unitary matrix $G(\bk)$, the sewing matrix transforms as $B_g (\bk_*) \to G (\bk_*)^{-1} \, B_g (\bk_*) \, G(\bk_*)$, which is a similarity transformation.
Consequently, its eigenvalues and trace are gauge invariant and can be computed independently of the specific choice of $G(\bk)$.

\section{Exciton formalism}
\label{app:exciton}

\subsection{Excitons and the exciton envelope wave function}
\label{app:ewf}
We begin by defining exciton states in a two-band system consisting of a conduction band $c$ and a valence band $v$.
We denote the corresponding creation and annihilation operators as
\bg
c^\dg_{\bk, c} = c^\dg_\bk,
\quad
c_{\bk, c} = c_\bk,
\quad
c^\dg_{\bk, v} = v^\dg_\bk,
\quad
c_{\bk, v} = v_\bk.
\eg
The ground state $\ket{\gs}$ is defined by filling the valence band:
\bg
\ket{{\rm GS}} = \prod_\bk \, v^\dg_\bk \ket{0}.
\eg
An exciton state with total momentum $\bp$ (measured relative to the ground state) is written as
\bg
\ket{\bp_{\rm exc}}
= \sum_\bk \, \phi_\bk (\bp) \, c^\dg_{\bk + \bp} v_\bk \ket{\gs}
\label{seq:exc_ket}
\eg
where $\phi_\bk (\bp)$ is the exciton envelope wave function (EWF).

In general, $\ket{\bp_{\rm exc}}$ is understood as an eigenstate of the full interacting Hamiltonian $\hat H$ in the single electron-hole excitation sector.
To analyze the structure of $\phi_\bk (\bp)$, it is convenient to introduce the projected Hamiltonian formalism, in which $\phi_\bk (\bp)$ appears as the $\bk$-component of an eigenvector.
We define the basis
\bg
\ket{\bk, \bp} = c^\dg_{\bk + \bp} v_\bk \ket{\gs},
\eg
so that $\ket{\bp_{\rm exc}} = \sum_\bk \, \phi_\bk (\bp) \ket{\bk, \bp}$.
Restricting the full Hamiltonian $\hat H$ to the single electron-hole excitation sector, the exciton eigenvalue problem is written as
\bg
\sum_{\bk'} \, \mc{H}_{\bk, \bk'} (\bp) \,\phi_{\bk'} (\bp) = E_{\rm exc} (\bp) \,\phi_{\bk} (\bp),
\quad
\mc{H}_{\bk,\bk'} (\bp) := \bra{\bk,\bp} \hat H \ket{\bk',\bp}.
\eg
Note that in translationally invariant systems, the projected Hamiltonian $\mc {H}$ is block diagonal in the total momentum, i.e., $\bra{\bk, \bp} \hat H \ket{\bk', \bp'} = 0$ unless $\bp = \bp' \pmod{\bG}$.
Thus, $\phi_\bk (\bp)$ is obtained as the $\bk$-component of an eigenvector of the projected Hamiltonian $\mc H(\bp)$.
In this notation, $\phi_\bk (\bp)$ is the $\bk$-component of the column vector $\phi (\bp)$, and $\mc{H}_{\bk, \bk'} (\bp)$ denotes the matrix element of $\mc{H} (\bp)$ in the $\bk$-th row and $\bk'$-th column.
The explicit construction of $\mc H_{\bk,\bk'} (\bp)$ and $\phi_\bk (\bp)$ from a microscopic model is detailed in Sec.~\ref{app:model}.

In general, for a given total momentum $\bp$, there can be multiple exciton eigenstates, corresponding to different eigenvalues $E_{\rm exc} (\bp)$ of the projected Hamiltonian.
In this work, we focus on a single exciton branch, typically the lowest-energy one, and suppress the corresponding band index for simplicity.
Unless otherwise stated, we assume that this exciton branch is isolated from others, i.e., separated by a finite gap.
Situations in which symmetry or topology enforces nodal structures are not considered here and are left for future study.

Since $\phi (\bp)$ is an eigenvector of the projected Hamiltonian $\mc H (\bp)$, we may impose the standard normalization
\bg
\sum_\bk \, \cm{\phi_\bk (\bp)} \phi_\bk (\bp) = 1.
\label{seq:exc_phi_norm}
\eg
Using the orthonormality of the basis states $\ket{\bk, \bp}$, $\brk{\bk, \bp} {\bk', \bp'} = \delta_{\bp, \bp'} \, \delta_{\bk, \bk'}$, one then finds
\bg
\brk{\bp_{\rm exc}}{\bp'_{\rm exc}}
= \delta_{\bp, \bp'} \sum_\bk \, \cm{\phi_\bk (\bp)} \phi_\bk (\bp)
= \delta_{\bp, \bp'}.
\label{seq:exc_state_norm}
\eg

Next, let us consider translations in the Brillouin zone by a reciprocal lattice vector $\bG$.
Using the periodic gauge for the Bloch states [Eq.~\eqref{seq:peri_gauge_2}], we have
\bg
c^\dg_{\bk + \bG} = c^\dg_\bk,
\quad
v_{\bk + \bG} = v_\bk,
\label{seq:peri_gauge_3}
\eg
which implies that the basis states satisfy
\bg
\ket{\bk + \bG, \bp} = \ket{\bk, \bp},
\quad
\ket{\bk, \bp + \bG} = \ket{\bk, \bp}.
\eg
As a result, the projected Hamiltonian obeys
\bg
\mc H_{\bk, \bk'}(\bp + \bG) = \mc H_{\bk, \bk'} (\bp).
\eg
Therefore, the eigenvectors at $\bp$ and $\bp + \bG$ can be simply related by a $U(1)$ phase, $\phi_\bk (\bp + \bG) = e^{i \zeta_\bp (\bG)} \phi_\bk (\bp)$.
We impose $\zeta_\bp (\bG)=0$, namely
\bg
\phi_\bk (\bp + \bG) = \phi_\bk (\bp),
\label{seq:exc_phi_periodic_1}
\eg
as the periodic gauge for the EWF.
Similarly, since crystal momentum is defined modulo a reciprocal lattice vector and the basis states satisfy $\ket{\bk + \bG, \bp} = \ket{\bk, \bp}$, the EWF is also periodic in $\bk$:
\bg
\phi_{\bk + \bG} (\bp) = \phi_\bk (\bp).
\label{seq:exc_phi_periodic_2}
\eg
In the periodic gauge for the band operators [Eq.~\eqref{seq:peri_gauge_3}] and the EWF [Eqs.~\eqref{seq:exc_phi_periodic_1} and \eqref{seq:exc_phi_periodic_2}], the exciton states $\ket{(\bp+\bG)_{\rm exc}}$ and $\ket{\bp_{\rm exc}}$ are identical:
\bg
\ket{(\bp+\bG)_{\rm exc}}
= \sum_\bk \, \phi_\bk (\bp+\bG) \, c^\dg_{\bk+\bp+\bG} v_\bk \ket{\gs}
= \sum_\bk \, \phi_\bk (\bp) \, c^\dg_{\bk+\bp} v_\bk \ket{\gs}
= \ket{\bp_{\rm exc}}.
\eg
Moreover, $\ket{\bp_{\rm exc}}$ is invariant under shifts of $\bk$ by a reciprocal lattice vector $\bG$.
To see this, we relabel $\bk \to \bk+\bG$ in Eq.~\eqref{seq:exc_ket}. In the periodic gauge, this does not change the state:
\bg
\ket{\bp_{\rm exc}} \to
\sum_\bk \, \phi_{\bk + \bG} (\bp) \, c^\dg_{\bk + \bG + \bp} v_{\bk + \bG} \ket{\gs}
= \sum_\bk \, \phi_\bk (\bp) \, c^\dg_{\bk + \bp} v_\bk \ket{\gs}
= \ket{\bp_{\rm exc}}.
\eg

Finally, we comment on gauge transformations of the EWF.
Under gauge transformations of the band operators,
\bg
c^\dg_\bk \to e^{i \theta_{c,\bk}} c^\dg_\bk, 
\quad
v^\dg_\bk \to e^{i \theta_{v,\bk}} v^\dg_\bk,
\eg
the basis state $\ket{\bk, \bp}$ transforms as $\ket{\bk, \bp} \to \ket{\bk, \bp} \, e^{i \theta_{c, \bk+\bp} - i \theta_{v, \bk} + i \theta_{\rm GS}}$, where $\theta_{\rm GS} = \sum_\bk \, \theta_{v, \bk}$.
Accordingly, the EWF transforms as
\bg
\phi_\bk (\bp) \to e^{-i \theta_{c,\bk+\bp} + i \theta_{v,\bk} - i \theta_{\rm GS}} \, \phi_\bk (\bp).
\eg
This transformation originates from a change of the Bloch basis and ensures that the exciton state $\ket{\bp}_{\rm exc}$ remains invariant~\cite{davenport2026exciton}.
In addition, one may perform a $U(1)$ gauge transformation
\bg
\phi_\bk (\bp) \to e^{i \theta_{\rm exc} (\bp)} \, \phi_\bk(\bp),
\label{seq:phi_gauge}
\eg
which reflects the intrinsic phase freedom of the eigenvector of the projected Hamiltonian $\mc H (\bp)$ and corresponds to a $U(1)$ phase rotation of the exciton state~\cite{davenport2026exciton}.
Since both transformations act multiplicatively by phase factors, the vanishing of $\phi_\bk (\bp)$ is invariant under these gauge choices.
Therefore, stable zeros of the EWF are gauge invariant.
We also note that analogous gauge-invariant zeros arise in flat-band wave functions~\cite{bergman2008band,rhim2019classification,hwang2021flat,hwang2021general}.

\subsection{Sewing matrices for exciton}
\label{app:exc_sewing}
To proceed, we define the exciton sewing matrix from the symmetry action on the exciton state.
We assume that both the single-particle and interaction parts of the Hamiltonian are separately symmetric under $g$, so that the symmetry action is well defined for $\ket{\bp_{\rm exc}}$.
Note that, since we consider a single exciton band, the exciton sewing matrix $\mc B_g(\bp)$ is a $U(1)$ phase factor.
Similarly, for nondegenerate conduction and valence bands, the band sewing matrices $B_{c,g}(\bk)$ and $B_{v,g}(\bk)$ are also $U(1)$ phase factors.

First, we note that the ground state transforms as
\bg
\hat g \ket{\gs} = \ket{\gs} \, B_{\gs, g} \, \mc K^{\tau_g},
\eg
where $B_{\gs, g} = \prod_\bk \, B_{v, g} (\bk)$, with $B_{v,g} (\bk)$ the sewing matrix of the valence ($v$) band at $\bk$ for $g$.
Then, we define the exciton sewing matrix $\mc B_g (\bp)$ through
\bg
\hat g \ket{\bp_{\rm exc}}
= \ket{(g\bp)_{\rm exc}} \, \mc B_g (\bp) \, B_{\gs, g} \, \mc K^{\tau_g}.
\label{seq:exc_sewing_def}
\eg
This means that $\mc B_g (\bp)$ is defined relative to the ground state.
This convention separates the symmetry transformation of the exciton excitation from that of the ground state, and leads to a simpler relation between symmetry and zero patterns.

Using the band sewing matrices $B_{c,g} (\bk)$ and $B_{v,g} (\bk)$ defined in Eq.~\eqref{seq:sym_cr_op_3}, the symmetry action on the basis state $\ket{\bk, \bp}$ follows directly.
This determines the transformation of the projected Hamiltonian $\mc H_{\bk, \bk'} (\bp)$ and the EWF $\phi_\bk (\bp)$.
At the level of the EWF, Eq.~\eqref{seq:exc_sewing_def} implies
\bg
B_{c,g} (\bk+\bp) \, B_{v,g} (\bk)^{-1} \, \cm{\phi_\bk (\bp)}^{\tau_g}
= \phi_{g\bk} (g\bp) \, \mc B_g(\bp).
\label{seq:exc_sewing_phi}
\eg
To derive Eq.~\eqref{seq:exc_sewing_phi}, we first note that
\ba
\hat{g} \ket{\bk, \bp}
=& (\hat{g} c^\dg_{\bk+\bp} \hat{g}^{-1}) \, (\hat{g} v_\bk \, \hat{g}^{-1}) \, \hat{g} \ket{\gs}
\nn
=& c^\dg_{g \bk + g \bp} B_{c,g} (\bk + \bp) \, v_{g \bk} \, B_{v,g} (\bk)^{-1} \, \ket{\gs} \, B_{\gs, g} \, \mc{K}^{\tau_g}
\nn
=& \ket{g \bk, g \bp} \, B_{c,g} (\bk + \bp) \, B_{v,g} (\bk)^{-1} \, B_{\gs, g} \, \mc{K}^{\tau_g}.
\label{seq:exc_basis_g}
\ea
Using this, we compute
\ba
\hat{g} \ket{\bp_{\rm exc}}
=& \sum_\bk \, \hat{g} \, \phi_\bk (\bp) \ket{\bk, \bp}
=\sum_\bk \, \cm{\phi_\bk (\bp)}^{\tau_g} \, \hat{g} \ket{\bk, \bp}
\nn
=& \sum_\bk \, \cm{\phi_\bk (\bp)}^{\tau_g} \, \ket{g \bk, g \bp} \, B_{c,g} (\bk + \bp) \, B_{v,g} (\bk)^{-1} \, B_{\gs, g} \, \mc{K}^{\tau_g}.
\label{seq:exc_sewing_deriv1}
\ea
On the other hand, $\ket{(g\bp)_{\rm exc}}$ can be written as
\bg
\ket{(g\bp)_{\rm exc}} = \sum_\bk \, \phi_{g\bk} (g\bp) \, \ket{g\bk, g\bp}.
\label{seq:exc_sewing_deriv2}
\eg
Comparing Eqs.~\eqref{seq:exc_sewing_deriv1} and \eqref{seq:exc_sewing_deriv2}, together with Eq.~\eqref{seq:exc_basis_g}, we obtain Eq.~\eqref{seq:exc_sewing_phi}.

We will discuss the stable zeros of the EWF in Sec.~\ref{app:zeros}.
For this purpose, Eq.~\eqref{seq:exc_sewing_phi} will be used extensively.
However, computing $\mc B_g (\bp)$ is also important, for example in numerical implementations.
The quantity $\mc B_g (\bp)$ can be computed as follows.
From Eq.~\eqref{seq:exc_sewing_phi}, by multiplying both sides by $\cm{\phi_{g\bk} (g\bp)}$ and using the normalization of $\phi_\bk (\bp)$, we obtain
\bg
\mc B_g (\bp)
= \sum_\bk \, \cm{\phi_{g\bk} (g\bp)} \, B_{c,g} (\bk+\bp) \, B_{v,g} (\bk)^{-1} \cm{\phi_\bk (\bp)}^{\tau_g}.
\label{eq:exc_sewing_formula}
\eg
In particular, for unitary $g$ ($\tau_g=0$), at a high-symmetry momentum $\bp_*$ satisfying $g \bp_* = \bp_* \pmod \bG$, $\mc B_g(\bp_*)$ reduces to the symmetry eigenvalue of the exciton at $\bp_*$.
In this case, $\mc B_g (\bp_*)$ is given by
\bg
\mc B_g (\bp_*) = \sum_\bk \, \cm{\phi_{g \bk} (\bp_*)} \, B_{c,g} (\bk + \bp_*) \, B_{v,g} (\bk)^{-1} \, \phi_\bk (\bp_*).
\label{seq:exc_sewing_hsm}
\eg
This expression is invariant under the $U(1)$ gauge transformation $\phi_\bk (\bp) \to e^{i \theta_{\rm exc} (\bp)} \phi_\bk (\bp)$ [Eq.~\eqref{seq:phi_gauge}], since the phase factors from $\cm{\phi_{g \bk} (\bp_*)}$ and $\phi_\bk (\bp_*)$ cancel.
Therefore, it provides a practical way to compute $\mc B_g (\bp_*)$ without gauge ambiguity.

Finally, we note the group-theoretical properties of the sewing matrices.
The sewing matrix satisfies the group relation
\bg
\mc B_{g_1 g_2} (\bp)
= \mc B_{g_1} (g_2 \bp) \,
\cm{\mc B_{g_2} (\bp)}^{\tau_{g_1}}.
\label{seq:sewing_rule}
\eg
This follows from
\ba
\hat{g}_1 \hat{g}_2 \, \ket{\bp_{\rm exc}}
=& \hat{g}_1 \, \ket{(g_2 \bp)_{\rm exc}} \, \mc B_{g_2}(\bp) \, \mc K^{\tau_{g_2}}
\nn
=& \ket{(g_1 g_2 \bp)_{\rm exc}} \, \mc B_{g_1} (g_2 \bp) \, \mc K^{\tau_{g_1}} \, \mc B_{g_2}(\bp) \, \mc K^{\tau_{g_2}}
\nn
=& \ket{(g_1 g_2 \bp)_{\rm exc}} \, \mc B_{g_1} (g_2 \bp) \, \cm{\mc B_{g_2} (\bp)}^{\tau_{g_1}} \, \mc K^{\tau_{g_1}+\tau_{g_2}},
\ea
which is to be compared with
\bg
(\hat{g}_1 \hat{g}_2) \, \ket{\bp_{\rm exc}}
= \ket{(g_1 g_2 \bp)_{\rm exc}} \, \mc B_{g_1 g_2} (\bp) \, \mc K^{\tau_{g_1 g_2}}.
\eg
Note that $\tau_{g_1 g_2} = \tau_{g_1} + \tau_{g_2} \pmod 2$.
In addition, $\mc B_g (\bp)$ is periodic in the Brillouin zone under the periodic gauge,
\bg
\mc B_g (\bp+\bG) = \mc B_g(\bp).
\label{seq:sewing_peri}
\eg

The symmetry action on the projected Hamiltonian $\mc H (\bp)$ can also be formulated by considering the transformation of the basis states $\ket{\bk,\bp}$ [Eq.~\eqref{seq:exc_basis_g}], which determines how symmetry operators act on $\mc H(\bp)$.
However, for the purposes of the present work, it is sufficient to focus on the exciton states $\ket{\bp_{\rm exc}}$ and the EWF $\phi_\bk(\bp)$.

\section{Exciton Wilson loops and Chern number}
\label{app:exc_wilson}
We now formulate the topological invariants of excitons using the Wilson-loop formalism~\cite{yu2011equivalent,fidkowski2011model,alexandradinata2014wilson}.
In particular, we consider the excitonic analogues of the Berry phase in one dimension (1D) and the Chern number in two dimensions (2D), which were discussed in the main text.
To this end, we first review the definition of exciton Berry connections and Wilson loops, following Ref.~\cite{davenport2026exciton}.
We then derive how exciton Wilson loops transform under symmetry operations.
Importantly, we show that the resulting transformation law takes the same form as that of electronic Wilson loops.
This allows us to directly relate excitonic topological invariants to symmetry eigenvalues, by adapting the well-established results established for electronic bands.

\subsection{Exciton Wilson loops and their symmetry transformation}
\label{app:exc_wilson_sym}
We begin by defining the exciton Berry connections.
Following Ref.~\cite{davenport2026exciton}, one can introduce two types of exciton Berry connections associated with the electron and hole sectors, respectively.
These are constructed by defining the position operator projected onto the conduction or valence band sector, and can be expressed in terms of the EWF $\phi_\bk (\bp)$ and the band Berry connections.
Here, we only state the resulting expressions and refer to Ref.~\cite{davenport2026exciton} for the detailed derivation.
Explicitly, the exciton Berry connections are given by
\bg
\mc A_c (\bp) = \sum_\bk \, i \cm{\phi_\bk (\bp)} \der_\bp \phi_\bk (\bp) + |\phi_\bk (\bp)|^2 A_c (\bk+\bp),
\nn
\mc A_v (\bp) = \sum_\bk \, i \cm{\phi_{\bk-\bp} (\bp)} \der_\bp \phi_{\bk-\bp} (\bp) + |\phi_{\bk-\bp} (\bp)|^2 A_v (\bk-\bp),
\eg
where $A_a (\bk)$ with $a=c,v$ denote the Berry connections of the conduction and valence bands.
In general, the two exciton Berry connections are not identical, reflecting the freedom in defining the exciton Berry phase, or equivalently, the exciton Wannier center~\cite{davenport2026exciton}.
This ambiguity originates from the fact that an exciton is a composite object of an electron and a hole, so that its center can be defined with respect to either constituent.

From these Berry connections $\mc A_a (\bp)$, one can define exciton Wilson lines.
In the continuum formulation, the exciton Wilson line along a path from $\bp$ to $\bp'$ is given by
\bg
\mc W_{a, \bp' \la \bp}
= \exp \left[ i \int_{\bp}^{\bp'} d \bq \, \mc A_a (\bq) \right].
\label{seq:wilson_def}
\eg
For practical purposes, it is convenient to work in a discretized formulation, in which the path is divided into segments.
Let $\{\bq_0=\bp, \bq_1, \dots, \bq_N=\bp'\}$ be a discretized path.
Then the Wilson line can be written as a product of line segments,
\bg
\mc W_{a, \bp' \la \bp}
= w_{a, \bq_N \la \bq_{N-1}} \, w_{a, \bq_{N-1} \la \bq_{N-2}} \cdots w_{a, \bq_2 \la \bq_1} \, w_{a, \bq_1 \la \bq_0} 
= \prod_{l=0}^{N-1} \, w_{a, \bq_{l+1} \la \bq_l},
\label{seq:wilson_line_def}
\eg
where each segment $w_{a, \bq_{l+1} \la \bq_l}$ is defined from overlaps of the EWF and band eigenstates.
Explicitly,
\bg
w_{c, \bq_{l+1} \la \bq_l} = \sum_\bk \, \brk{u_{\bk + \bq_{l+1}, c}}{u_{\bk + \bq_l, c}} \, \cm{\phi_\bk ({\bq_{l+1}})} \phi _\bk (\bq_l),
\nn
w_{v, \bq_{l+1} \la \bq_l} = \sum_\bk \, \brk{u_{\bk + \bq_{l+1} - \bq_l, v}}{u_{\bk, v}} \, \cm{\phi_\bk (\bq_{l+1})} \phi_{\bk + \bq_{l+1} - \bq_l} (\bq_l).
\eg
The discretized formulation is particularly convenient for deriving the symmetry transformation of Wilson lines segment by segment: for electronic Wilson loops, symmetry transformations in the discretized formulation have been discussed, for example, in Ref.~\cite{alexandradinata2016topological}, while derivations in the continuum formulation can be found in Ref.~\cite{hwang2019fragile}.

When the path is closed, namely when $\bp' = \bp$ up to a reciprocal lattice vector $\bG$, the Wilson line is called a Wilson loop.
The Wilson-loop formalism provides a natural framework for defining topological invariants of excitons.
In 1D, the exciton Berry phase is given by the phase of a Wilson loop evaluated along a closed path in momentum space, for example along the Brillouin zone from $p=0$ to $p=2\pi$.
In general, the Berry phases computed from $\mc A_c (\bp)$ and $\mc A_v (\bp)$ differ~\cite{davenport2026exciton} in the absence of symmetry such as inversion.
In 2D, the exciton Chern number can be obtained from the winding of the Wilson-loop spectrum as a function of momentum, as in the case of electronic bands~\cite{yu2011equivalent,alexandradinata2014wilson}.
However, the Chern number obtained from the Wilson loops is the same for $a=c$ and $a=v$~\cite{davenport2026exciton}.

To analyze symmetry constraints, it is sufficient to consider the transformation of individual Wilson-line segments in the discretized formulation.
The transformation of the full Wilson line then follows by taking the product over segments.
In the following, we derive the symmetry transformation of the Wilson line by explicitly evaluating how each segment transforms under a symmetry operation $g$.
Let us first consider the case with $a=c$.
We aim to relate $w_{c, g (\bq + \delta \bq) \la g \bq}$ with $w_{c, \bq + \delta \bq \la \bq}$, where $\delta \bq$ is a small momentum increment along the path:
\ba
w_{c, g (\bq + \delta \bq) \la g \bq}
=& \sum_\bk \, \brk{u_{\bk +g \bq + g \delta \bq, c}}{u_{\bk + g \bq, c}} \, \cm{\phi_\bk (g \bq + g\delta \bq)} \, \phi_\bk (g \bq)
\nn
=& \sum_\bk \, \brk{u_{g \bk +g \bq + g \delta \bq, c}}{u_{g \bk + g \bq, c}} \, \cm{\phi_{g \bk}(g \bq + g \delta \bq)} \, \phi_{g \bk}(g \bq)
\ea
In the summation, the overlap term can be rewritten as
\ba
\brk{u_{g \bk +g \bq + g \delta \bq, c}}{u_{g \bk + g \bq, c}}
=& B_{c,g} (\bk + \bq + \delta \bq) \, \cm{\bra{u_{\bk + \bq + \delta \bq, c}}}^{\tau_g} \, U_g^{-1} \, e^{i g (\bk+ \bq + \delta \bq) \cdot \bb v_g}
\nn
& \times e^{- i g (\bk + \bq) \cdot \bb v_g} \, U_g \, \cm{\ket{u_{\bk+ \bq, c}}}^{\tau_g} \, B_{c, g} (\bk + \bq)^{-1},
\ea
where we used Eq.~\eqref{seq:sewing_u}.
The EWF term can be expressed, using Eq.~\eqref{seq:exc_sewing_phi}, as
\ba
\cm{\phi_{g \bk} (g \bq + g \delta \bq)} \, \phi_{g \bk}(g \bq)
=& \mc B_g (\bq + \delta \bq) \, \cm{\phi_\bk(\bq + \delta \bq)}^{\tau_g + 1} \, B_{c, g} (\bk + \bq + \delta \bq)^{-1} \, B_{c, g} (\bk + \bq) \, \cm{\phi_\bk (\bq)}^{\tau_g} \, \mc B_g (\bq)^{-1}.
\ea
Combining these expressions, we obtain
\ba
w_{c, g (\bq + \delta \bq) \la g \bq}
=& e^{i g \delta \bq \cdot \bb v_g} \, \sum_\bk \, \mc B_g (\bq + \delta \bq) \, \mc K^{\tau_g} \, \brk{u_{\bk + \bq + \delta \bq, c}}{u_{\bk + \bq, c}} \, \cm{\phi_\bk (\bq + \delta \bq)} \, \phi_\bk (\bq) \, \mc K^{\tau_g} \, \mc B_g (\bq)^{-1}
\nn
=& e^{i g \delta \bq \cdot \bb v_g} \, \mc B_g (\bq + \delta \bq) \, \cm{w_{c, \bq + \delta \bq \la \bq}}^{\tau_g} \, \mc B_g (\bq)^{-1}.
\label{seq:wilson_segment_c}
\ea
The case of $a=v$ can be obtained in the same manner, yielding
\bg
w_{v, g (\bq + \delta \bq) \la g \bq}
= e^{i g \delta \bq \cdot \bb v_g} \, \mc B_g (\bq + \delta \bq) \, \cm{w_{v, \bq + \delta \bq \la \bq}}^{\tau_g} \, \mc B_g (\bq)^{-1}.
\label{seq:wilson_segment_v}
\eg
By substituting Eqs.~\eqref{seq:wilson_segment_c} and \eqref{seq:wilson_segment_v} into Eq.~\eqref{seq:wilson_line_def}, we obtain
\bg
\mc W_{a, g\bp' \la g\bp} = e^{i g (\bp'-\bp) \cdot \bb v_g} \, \mc B_g (\bp') \, \cm{\mc W_{a, \bp' \la \bp}}^{\tau_g} \, \mc B_g (\bp)^{-1},
\label{seq:wilson_sym}
\eg
for both $a=c,v$.
Here, it is understood that the path defining $\mc W_{a, \bp' \la \bp}$ is mapped to that of $\mc W_{a, g\bp' \la g\bp}$ under the symmetry action $g$.
The same constraint is imposed on both $\mc W_{c, \bp' \la \bp}$ and $\mc W_{v, \bp' \la \bp}$ by the exciton sewing matrix $\mc B_g$.

\subsection{Symmetry constraints on exciton Berry phase and Chern number}
\label{app:exc_chern}
Above, we showed that the symmetry transformation law of exciton Wilson loops takes the same form for both $a=c$ and $v$, as expressed in Eq.~\eqref{seq:wilson_sym}.
Moreover, Eq.~\eqref{seq:wilson_sym} takes the same form as the symmetry transformation of electronic Wilson loops~\cite{alexandradinata2016topological}.
Therefore, the determination of exciton topological invariants from symmetry eigenvalues follows directly from the corresponding Wilson-loop results for electronic bands, including the Berry phase in 1D inversion-symmetric systems~\cite{alexandradinata2014wilson,davenport2026exciton} and the Chern number in 2D rotation-symmetric systems~\cite{fang2012bulk}.
For completeness, we explicitly work out these cases below.

\tocless{\subsubsection{1D Berry phase and inversion symmetry}}{}
We first consider a 1D system with inversion symmetry $P$.
The Brillouin zone has high-symmetry momenta (HSM) at $p_*=0$ and $\pi$, which satisfy $P p_* = -p_* = p_* \pmod{2\pi}$.
Since $P^2=1$, the inversion sewing matrix $\mc B_P (p_*)$ takes values $\pm 1$ at the HSM $p_*$.
The exciton Berry phase $\gamma_a$ ($a=c,v$) is defined as the phase of the (abelian) Wilson loop over the Brillouin zone,
\bg
\gamma_a = \arg \, \mc W_{a, \pi \la -\pi} \, \in [0, 2\pi).
\eg
The Wilson loop can be divided into two segments as
\bg
\mc W_{a, \pi \la -\pi} = \mc W_{a, \pi \la 0} \, \mc W_{a, 0 \la -\pi}.
\label{seq:1d_wilson}
\eg
First, note that the inverse of $\mc W_{a, 0 \la -\pi}$ is $\mc W_{a, -\pi \la 0}$, corresponding to reversing the path direction.
We now relate $\mc W_{a, -\pi \la 0}$ to $\mc W_{a, \pi \la 0}$ using the symmetry transformation in Eq.~\eqref{seq:wilson_sym}.
\bg
 \mc W_{a, -\pi \la 0} = \mc B_P (\pi) \, \mc W_{a, \pi \la 0} \,\mc B_P (0)^{-1}.
\eg
Thus, Eq.~\eqref{seq:1d_wilson} becomes
\bg
\mc W_{a, \pi \la -\pi}
= \mc W_{a, \pi \la 0} \, \mc W_{a, -\pi \la 0}^{-1}
= \mc W_{a, \pi \la 0} \, \mc B_P (0) \, \mc W_{a, \pi \la 0}^{-1} \,\mc B_P (\pi)^{-1}
= \mc B_P (0) \,\mc B_P (\pi)^{-1}.
\label{seq:1d_berry_P}
\eg
Hence, the exciton Berry phase $\gamma_a$ is completely determined by the inversion eigenvalues at the HSM: when the inversion eigenvalues at $0$ and $\pi$ are the same (opposite), $\gamma_a = 0$ ($\pi$).
Moreover, in the presence of inversion symmetry, $\gamma_c = \gamma_v$, unlike the case without a quantizing symmetry.
Based on this equivalence, in the main text we defined the exciton Wannier center as
\bg
x_{\rm exc} = \frac{\gamma_c}{2\pi} = \frac{\gamma_v}{2\pi} \pmod 1,
\quad
x_{\rm exc} \in \{0, 1/2\}.
\label{seq:wannier_def}
\eg

\begin{figure*}[t]
\centering
\includegraphics[width=0.65\textwidth]{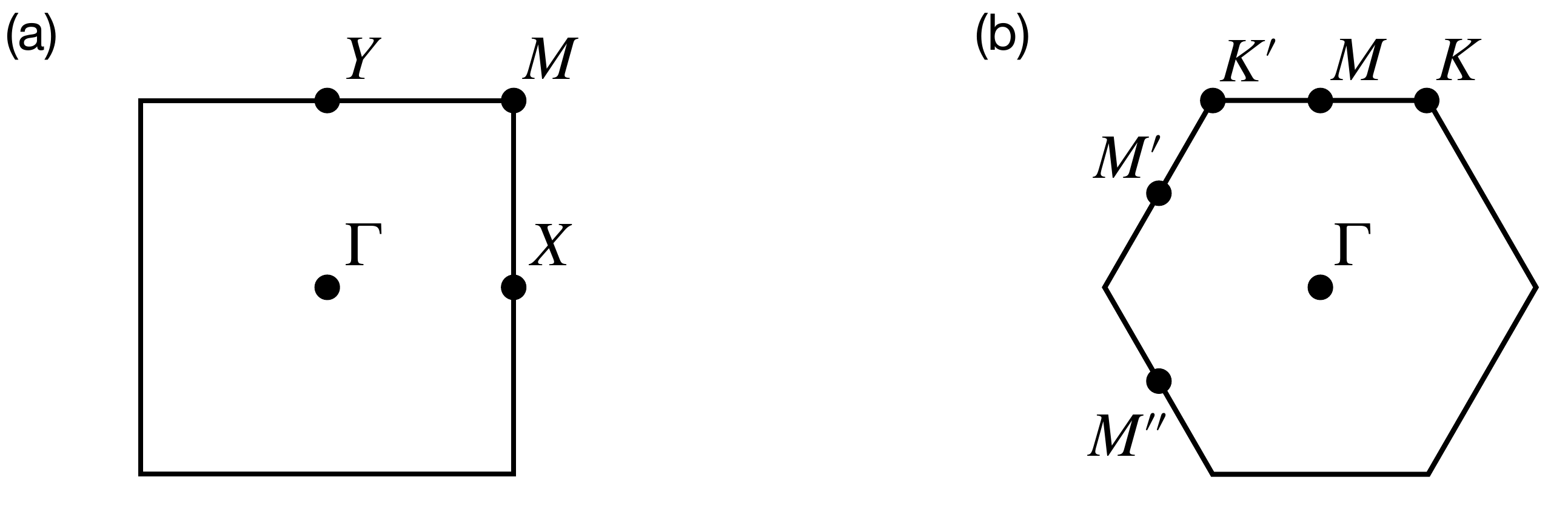}
\caption{High-symmetry momenta (HSM) in the Brillouin zones of $C_n$-symmetric systems.
(a) Square Brillouin zone relevant for $C_2$ and $C_4$ symmetries, with reciprocal lattice vectors $\bb g_1 = (2\pi,0)$ and $\bb g_2 = (0,2\pi)$.
The HSM are $\Gamma=(0,0)$, $X=(\pi,0)$, $Y=(0,\pi)$, and $M=(\pi,\pi)$.
(b) Hexagonal Brillouin zone relevant for $C_3$ and $C_6$ symmetries, with reciprocal lattice vectors $\bb g_1 = \left( 2\pi,-\frac{2\pi}{\sqrt 3} \right)$ and $\bb g_2 = \left( 0,\frac{4\pi}{\sqrt 3} \right)$.
The HSM are $\Gamma=(0,0)$, $K = \left( \frac{2\pi}{3}, \frac{2\pi}{\sqrt 3} \right)$, $K' = \left( -\frac{2\pi}{3}, \frac{2\pi}{\sqrt 3} \right)$, and $M = (0, \frac{2\pi}{\sqrt 3})$, together with symmetry-related points $M' = (-\pi, \frac{\pi}{\sqrt 3})$ and $M'' = (-\pi, -\frac{\pi}{\sqrt 3})$.
For $C_3$, the rotation-invariant HSM are $\{\Gamma, K, K'\}$.
For $C_6$, all six points are invariant under a nontrivial subgroup: $\Gamma$ under $C_6$, $K, K'$ under $C_3$, and $M, M', M''$ under $C_2$.
Symmetry relations include $K'=C_6 K$, $M'=C_6 M$, and $M''=C_3 M$.}
\label{sfig:2D_BZ}
\end{figure*}

\tocless{\subsubsection{2D Chern number and rotation symmetry}}{}
We now consider a two-dimensional system with $C_n$ rotation symmetry ($n=2,3,4,6$).
In this case, the Brillouin zone contains HSM that are invariant under $C_m$ rotations with $m \le n$.
These momenta, as well as the corresponding Brillouin-zone geometry for each $C_n$, are illustrated in Fig.~\ref{sfig:2D_BZ}.
As shown in Ref.~\cite{fang2012bulk}, the Chern number $\mc C$ modulo $n$ is determined by the rotation eigenvalues at these HSM.
This result can be understood naturally within the Wilson-loop formalism.
The key idea is to decompose an appropriate Wilson loop into segments and analyze how these segments are related by symmetry through the sewing matrices~\cite{fang2012bulk}.
By systematically applying the symmetry constraints derived in Eq.~\eqref{seq:wilson_sym}, one can determine the mod-$n$ Chern number (i.e., $\mc C \mod n$) from the symmetry eigenvalues.
For the reader's convenience, we present explicit derivations for the $C_2$ and $C_4$ cases below, while the $C_3$ and $C_6$ cases are treated by outlining the setup and presenting the final expressions.
For more detailed discussions, we refer to Ref.~\cite{fang2012bulk}, and to Ref.~\cite{alexandradinata2016berry} for broader discussions of Wilson loops in the presence of $C_n$ symmetries.
In what follows, we restrict to the spinless case, so that $(C_n)^n = 1$.

\paragraph{$C_2$ symmetry:}
We first consider the case of $C_2$ rotation symmetry.
In this case, the Brillouin zone contains four HSM, $\Gamma=(0,0)$, $X=(\pi,0)$, $Y=(0,\pi)$, and $M=(\pi,\pi)$, which satisfy $C_2 \bp_* = -\bp_* = \bp_* \pmod \bG$.
We take the reciprocal lattice vectors as $\bb g_1 = (2\pi,0)$ and $\bb g_2 = (0,2\pi)$.

We consider the closed path $\mc L: X - \bb g_1 \to \Gamma \to X \to M \to Y \to M - \bb g_1 \to X - \bb g_1$, where each segment is taken to be a straight line in the Brillouin zone.
The Wilson loop along $\mc L$ is given by
\bg
\mc W_{\mc L} = \exp \left[ i \oint_{\mc L} d\bp \cdot \mc A (\bp) \right]
= \exp \left[ i \int_{\mc S} d^2 p \, \mc F_{xy}(\bp) \right],
\eg
where the second equality follows from Stokes' theorem, and $\mc S$ denotes a surface bounded by $\mc L$ (In the periodic gauge, the Berry connection may not be smooth, but the Wilson loop remains well defined and satisfies Stokes' relation at the exponentiated level.
We adopt this gauge since it simplifies the treatment of sewing matrices and Wilson loops).
Here, we suppress the index $a=c,v$, since the exciton Chern number $\mc C_{\rm exc}$ obtained from $\mc A_c$ and $\mc A_v$ is identical~\cite{davenport2026exciton}, independent of symmetry.
For our purposes, both choices lead to the same symmetry constraints on the Wilson loops, and therefore yield the same expression in terms of sewing matrices, which in turn determines the same value of $\mc C_{\rm exc}$ modulo $n$.
This convention will be adopted throughout the discussion of rotation symmetries below.

Due to $C_2$ symmetry, the integral over $\mc S$ equals half of the total Berry curvature integral over the Brillouin zone, and hence
\bg
\mc W_{\mc L} = e^{i \pi \mc C_{\rm exc}}.
\eg
We now evaluate $\mc W_{\mc L}$ by decomposing it into Wilson line segments,
\bg
\mc W_{\mc L}
= \mc W_{X - \bb g_1 \la M - \bb g_1} \, \mc W_{M - \bb g_1 \la Y \la M} \, \mc W_{M \la X} \, \mc W_{X \la \Gamma \la X - \bb g_1}.
\eg
Under the periodic gauge, we have $\mc W_{X - \bb g_1 \la M - \bb g_1} = \mc W_{X \la M}$, and therefore
\bg
\mc W_{X - \bb g_1 \la M - \bb g_1} \, \mc W_{M \la X}
= \mc W_{X \la M} \, \mc W_{M \la X}
= \mc W_{M \la X}^{-1} \, \mc W_{M \la X}
= 1.
\eg
The remaining segments can be evaluated using the same method as in the 1D inversion case [Eq.~\eqref{seq:1d_wilson}].
In particular, one finds $\mc W_{X \la \Gamma \la X - \bb g_1} = \mc B_{C_2}(\Gamma) \, \mc B_{C_2}(X)^{-1}$ and $\mc W_{M - \bb g_1 \la Y \la M} = \mc B_{C_2}(Y) \, \mc B_{C_2}(M)^{-1}$.
Combining these contributions, we obtain
\bg
\mc W_{\mc L}
= \mc B_{C_2}(\Gamma) \, \mc B_{C_2}(X)^{-1} \, \mc B_{C_2}(Y) \, \mc B_{C_2}(M)^{-1}.
\eg
Since $\mc B_{C_2}(\bp_*) = \pm 1$ at the HSM, this simplifies to
\bg
e^{i \pi \mc C_{\rm exc}}
= \mc B_{C_2}(\Gamma) \, \mc B_{C_2}(X) \, \mc B_{C_2}(Y) \, \mc B_{C_2}(M).
\eg
Thus, the exciton Chern number modulo $2$ is determined by the $C_2$ eigenvalues at the HSM.

\paragraph{$C_4$ symmetry:}
We next consider the case of $C_4$ rotation symmetry.
In this case, the HSM are $\Gamma = (0,0)$, $M=(\pi,\pi)$, $X=(\pi,0)$, and $Y=(0,\pi)$.
Unlike the $C_2$ case, $Y$ is not an independent HSM, since $Y = C_4 X$.
We take the reciprocal lattice vectors as $\bb g_1 = (2\pi,0)$ and $\bb g_2 = (0,2\pi)$, under which $C_4$ acts as $\bb g_1 \to \bb g_2$ and $\bb g_2 \to -\bb g_1$.

We consider the closed path $\mc L: \Gamma \to X \to M \to Y \to \Gamma$.
As in the $C_2$ case, $\mc W_{\mc L}$ is given by the exponentiated Berry-curvature integral over a surface $\mc S$ bounded by $\mc L$.
Since $\mc S$ covers one quarter of the Brillouin zone and the Berry curvature is related by $C_4$ symmetry, we obtain
\bg
\mc W_{\mc L} = e^{i \frac{\pi}{2} \mc C_{\rm exc}}.
\eg

We now evaluate $\mc W_{\mc L}$ in terms of sewing matrices.
Writing the loop as a product of Wilson lines, we have
\bg
\mc W_{\mc L}
= \mc W_{\Gamma \la Y} \, \mc W_{Y \la M} \, \mc W_{M \la X} \, \mc W_{X \la \Gamma}.
\label{seq:c4_wilson_part0}
\eg
We first relate $\mc W_{Y \la \Gamma}$ to $\mc W_{X \la \Gamma}$.
Using Eq.~\eqref{seq:wilson_sym} with $g=C_4$, one finds $\mc W_{Y \la \Gamma} = \mc B_{C_4}(X) \, \mc W_{X \la \Gamma} \, \mc B_{C_4}(\Gamma)^{-1}$.
Here, $\Gamma$ is a $C_4$-invariant momentum, so $\mc B_{C_4}(\Gamma)$ is the gauge-invariant $C_4$ eigenvalue at $\Gamma$.
By contrast, $X$ is not invariant under $C_4$, and therefore $\mc B_{C_4}(X)$ should not be interpreted as a symmetry eigenvalue.
Using $\mc W_{\Gamma \la Y} = \mc W_{Y \la \Gamma}^{-1}$, we obtain
\bg
\mc W_{\Gamma \la Y} \, \mc W_{X \la \Gamma}
= \mc W_{Y \la \Gamma}^{-1} \, \mc W_{X \la \Gamma}
= \mc B_{C_4}(\Gamma) \, \mc B_{C_4}(X)^{-1}.
\label{seq:c4_wilson_part1}
\eg
Next, we consider the remaining pair $\mc W_{Y \la M}$ and $\mc W_{M \la X}$.
We derive two relations for $\mc W_{M' \la Y}$, where $M' = M - \bb g_1$.
First, using $C_4$, we have $\mc W_{M' \la Y} = \mc B_{C_4}(M) \, \mc W_{M \la X} \, \mc B_{C_4}(X)^{-1}$.
Second, using $C_2$, we have $\mc W_{M' \la Y} = \mc B_{C_2}(M) \, \mc W_{M \la Y} \, \mc B_{C_2}(Y)^{-1}$.
Comparing these expressions, we obtain
\bg
\mc W_{Y \la M} \mc W_{M \la X}
= \mc W_{M \la Y}^{-1} \mc W_{M \la X}
= \mc B_{C_2}(M) \, \mc B_{C_2}(Y)^{-1} \, \mc B_{C_4}(X) \, \mc B_{C_4}(M)^{-1}.
\label{seq:c4_wilson_part2}
\eg
Therefore, combining Eqs.~\eqref{seq:c4_wilson_part1} and \eqref{seq:c4_wilson_part2}, Eq.~\eqref{seq:c4_wilson_part0} becomes
\bg
\mc W_{\mc L}
= ( \mc W_{\Gamma \la Y} \, \mc W_{X \la \Gamma} ) ( \mc W_{Y \la M} \, \mc W_{M \la X} )
= \mc B_{C_4}(\Gamma) \, \mc B_{C_2}(M) \, \mc B_{C_4}(M)^{-1} \, \mc B_{C_2}(Y)^{-1}.
\label{seq:C4_loop_intermediate}
\eg

We now simplify this expression using the group relation of sewing matrices [Eq.~\eqref{seq:sewing_rule}] together with their periodicity in the Brillouin zone [Eq.~\eqref{seq:sewing_peri}].
First, at $M$, we have
\bg
\mc B_{C_2}(M) = \mc B_{C_4 C_4}(M) = \mc B_{C_4}(C_4 M) \, \mc B_{C_4}(M) = \mc B_{C_4}(M)^2,
\eg
since $C_4 M = M \pmod \bG$.
Second, at $Y=C_4 X$, we use the group relation to obtain
\ba
\mc B_{C_2}(Y)
=& \mc B_{C_2}(Y) \mc B_{C_4} (X) \mc B_{C_4} (X)^{-1}
= [ \mc B_{C_2}(C_4 X) \mc B_{C_4} (X) ] \mc B_{C_4} (X)^{-1}
= \mc B_{C_2 C_4}(X) \, \mc B_{C_4}(X)^{-1}
\nn
=& \mc B_{C_4}(C_2 X) \, \mc B_{C_2}(X) \, \mc B_{C_4}(X)^{-1}
= \mc B_{C_2 C_4}(X) \, \mc B_{C_4}(X)^{-1}
= \mc B_{C_4}(X) \, \mc B_{C_2}(X) \, \mc B_{C_4}(X)^{-1}
= \mc B_{C_2}(X),
\ea
as expected, since $X$ and $Y$ are related by $C_4$ symmetry.
Finally, since $\mc B_{C_2}(X) = \pm 1$, we have $\mc B_{C_2} (X)^{-1}=\mc B_{C_2} (X)$.
Substituting these relations into Eq.~\eqref{seq:C4_loop_intermediate}, we find
\bg
e^{i \frac{\pi}{2} \mc C_{\rm exc}}
= \mc B_{C_4}(\Gamma) \, \mc B_{C_4}(M) \, \mc B_{C_2}(X).
\eg
This expression can be written in the compact form
\bg
e^{i \frac{\pi}{2} \mc C_{\rm exc}}
= \prod_{\bp_*} \mc B_{g_{\bp_*}}(\bp_*),
\eg
where $\bp_* \in \{\Gamma,X,M\}$, and $g_{\bp_*}$ denotes the highest rotation symmetry leaving $\bp_*$ invariant.
Thus, $g_\Gamma=g_M=C_4$, while $g_X=C_2$.

\paragraph{$C_3$ symmetry:}
We next consider $C_3$ rotation symmetry.
The HSM are $\Gamma$, $K$, and $K'$, given by $\Gamma=(0,0)$, $K=\left(\frac{2\pi}{3}, \frac{2\pi}{\sqrt 3}\right)$, and $K'=\left(-\frac{2\pi}{3}, \frac{2\pi}{\sqrt 3}\right)$, all of which are invariant under $C_3$ modulo reciprocal lattice vectors.
We take the reciprocal lattice vectors as $\bb g_1 = \left(2\pi, -\frac{2\pi}{\sqrt 3}\right)$ and $\bb g_2 = \left(0, \frac{4\pi}{\sqrt 3}\right)$, under which $C_3$ acts as $\bb g_1 \to \bb g_2$ and $\bb g_2 \to -\bb g_1 - \bb g_2$.

We consider the closed path $\mc L: \Gamma \to K' + \bb g_1 \to K \to K' \to \Gamma$.
The Wilson loop $\mc W_{\mc L}$ is given by the exponentiated Berry-curvature integral over a surface bounded by $\mc L$, which by $C_3$ symmetry yields
\bg
\mc W_{\mc L} = e^{i \frac{2\pi}{3} \mc C_{\rm exc}}.
\eg
By decomposing $\mc W_{\mc L}$ into Wilson-line segments and applying the symmetry constraints from Eq.~\eqref{seq:wilson_sym}, one finds $e^{i \frac{2\pi}{3} \mc C_{\rm exc}} = \mc B_{C_3}(\Gamma) \, \mc B_{C_3}(K) \, \mc B_{C_3}(K')^{-2}$.
Using the group structure of sewing matrices [Eq.~\eqref{seq:sewing_rule}], this simplifies to
\bg
e^{i \frac{2\pi}{3} \mc C_{\rm exc}}
= \mc B_{C_3}(\Gamma) \, \mc B_{C_3}(K) \, \mc B_{C_3}(K').
\eg
Thus, the exciton Chern number modulo $3$ is determined by the $C_3$ eigenvalues at the HSM.

\paragraph{$C_6$ symmetry:}
We now consider $C_6$ rotation symmetry.
The HSM are $\Gamma$, $K$, and $M$, given by $\Gamma=(0,0)$, $K=\left(\frac{2\pi}{3}, \frac{2\pi}{\sqrt 3}\right)$, and $M=\left(0, \frac{2\pi}{\sqrt 3}\right)$, which are invariant under $C_6$, $C_3$, and $C_2$, respectively.
We take the reciprocal lattice vectors as $\bb g_1 = \left(2\pi, -\frac{2\pi}{\sqrt 3}\right)$ and $\bb g_2 = \left(0, \frac{4\pi}{\sqrt 3}\right)$, under which $C_6$ acts as $\bb g_1 \to \bb g_1 + \bb g_2$ and $\bb g_2 \to -\bb g_1$.

We consider the closed path $\mc L: \Gamma \to K \to M \to K' \to \Gamma$, where $K' = C_6 K$.
As before, $\mc W_{\mc L}$ is given by the exponentiated Berry-curvature integral, which yields
\bg
\mc W_{\mc L} = e^{i \frac{\pi}{3} \mc C_{\rm exc}}.
\eg
By applying both $C_6$ and $C_2$ symmetry constraints to the Wilson-line segments, one finds
\bg
e^{i \frac{\pi}{3} \mc C_{\rm exc}}
= \mc B_{C_6} (\Gamma) \, \mc B_{C_2} (K) \, \mc B_{C_6} (K)^{-1} \, \mc B_{C_2}(M),
\eg
which simplifies to
\bg
e^{i \frac{\pi}{3} \mc C_{\rm exc}}
= \mc B_{C_6} (\Gamma) \, \mc B_{C_3} (K) \, \mc B_{C_2} (M).
\eg
Thus, the exciton Chern number modulo $6$ is determined by the highest rotation eigenvalues at the HSM.

\paragraph{Summary and remarks:}
We summarize the expressions for the exciton Chern number for $C_n$ with $n=2,3,4,6$.
For $C_n$ symmetry, the exciton Chern number modulo $n$ can be expressed as
\bg
e^{i \frac{2\pi}{n} \mc C_{\rm exc}}
= \prod_{\bp_* \in {\rm mHSM}} \mc B_{g_{\bp_*}}(\bp_*),
\label{seq:chern_formula}
\eg
where the product runs over a minimal set of HSM ($\rm mHSM$), containing one representative from each orbit under $C_n$.
For example, for $C_6$ symmetry, the mHSM is $\{\Gamma, K, M\}$, while for $C_4$ it is $\{\Gamma, X, M\}$.
Also, $g_{\bp_*}$ denotes the highest rotation symmetry that leaves $\bp_*$ invariant.
A subtle point concerns the definition of the exciton sewing matrix in Eq.~\eqref{seq:exc_sewing_def}, where the ground-state contribution $B_{\gs, g}$ is separated out explicitly.
One may alternatively define the sewing matrix without this factor.
However, this redefinition does not affect the final result for $e^{i \frac{2\pi}{n} \mc C_{\rm exc}}$.
To see this, note that each sewing matrix at HSM acquires an extra factor $B_{\gs, g_{\bp_*}}$.
The total contribution is therefore $\prod_{\bp_* \in {\rm HSM}} B_{\gs,g_{\bp_*}}$, which equals 1 for all $n$.
For example, in the $C_6$ case, the extra factors combine as
\bg
B_{\gs,C_6} \, B_{\gs,C_3} \, B_{\gs,C_2}
= B_{\gs, C_6 C_3 C_2} = B_{\gs,1} = 1,
\eg
which explicitly shows the cancellation.

\section{Structure of stable-zero patterns}
\label{app:zeros}
In this section, we analyze the structure of stable zero patterns of the exciton EWF under inversion symmetry in 1D and $C_n$ rotation symmetry in 2D.
Rather than attempting an exhaustive enumeration, which becomes impractical in 2D, we describe a systematic procedure for determining the allowed zero patterns from symmetry.
In 1D with inversion symmetry, the problem is sufficiently simple that all possible zero patterns can be fully listed.
In 2D with $C_n$ rotation symmetry, the number of possibilities grows rapidly, and we instead describe their general structure and list representative patterns of physical interest.
Finally, we restrict the zero patterns to $\bp=0$, corresponding to excitons with zero total momentum.
In this case, we summarize the stable zero patterns and their relation to the relative band Chern number modulo $n$.

\subsection{Inversion symmetry in one dimension}
\label{app:zero_1d}
We first consider a 1D system with inversion symmetry $P$.
The HSM are $k_*=0$ and $\pi$, which satisfy $P k_* = -k_* = k_* \pmod{2\pi}$.
At these momenta, the exciton sewing matrix $\mc B_P(k_*)$ reduces to the inversion eigenvalue, taking values $\pm 1$.
The symmetry constraint Eq.~\eqref{seq:exc_sewing_phi} at $k_*$ becomes
\bg
\left[ B_{c,P} (k_* + p_*) \, B_{v,P} (k_*)^{-1} \, \mc B_P(p_*)^{-1} \right] \phi_{k_*} (p_*)
= \phi_{k_*} (p_*).
\eg
The prefactor in brackets takes either $+1$ or $-1$.
If it is $+1$, the above equation becomes an identity, while if it is $-1$, it enforces $\phi_{k_*} (p_*)=0$, yielding a stable zero.
We now relate the stable zero patterns to the Wannier-center shifts
\bg
s_{c/v} = x_{c/v} - x_{\rm exc} \pmod 1.
\label{seq:1d_shift_def}
\eg
As discussed in Sec.~\ref{app:exc_chern}, the Wannier centers are determined by inversion eigenvalues at the HSM $0$ and $\pi$: they are $0 \pmod 1$ if the eigenvalues coincide, and $1/2 \pmod 1$ otherwise [See also Eqs.~\eqref{seq:1d_berry_P} and \eqref{seq:wannier_def}].

Following the notation in the main text, let us write $\xi_c(k) := B_{c,P} (k)$, $\xi_v(k) := B_{v,P} (k)$, and $\xi_{\rm exc}(p) := \mc B_P (p)$.
At $p_*=0$ and $\pi$, we obtain
\bg
\left[ \xi_c (k_*) \, \xi_v (k_*)^{-1} \, \xi_{\rm exc} (0)^{-1} \right] \, \phi_{k_*} (0) = \phi_{k_*} (0),
\quad
\left[ \xi_c (k_* + \pi) \, \xi_v (k_*)^{-1} \, \xi_{\rm exc} (\pi)^{-1} \right] \, \phi_{k_*} (\pi) = \phi_{k_*} (\pi),
\label{seq:inv_constraint1}
\eg
These are the same constraints used in the main text.
We now spell out the remaining cases explicitly.

From Eq.~\eqref{seq:inv_constraint1}, $\phi_{k_*} (0)$ is a stable zero if and only if $\xi_c (k_*) \, \xi_v (k_*)^{-1} \, \xi_{\rm exc} (0)^{-1} = -1$, and similarly $\phi_{k_*} (\pi)$ is a stable zero if and only if $\xi_c (k_* + \pi) \, \xi_v (k_*)^{-1} \, \xi_{\rm exc} (\pi)^{-1} = -1$.
As discussed in the main text, if both $\phi_{k_*} (0)$ and $\phi_{k_*} (\pi)$ are stable zeros, or if both are nonzero, then
\bg
\xi_c (k_*) \, \xi_v (k_*)^{-1} \, \xi_{\rm exc} (0)^{-1}  = \xi_c (k_* + \pi) \, \xi_v (k_*)^{-1} \, \xi_{\rm exc} (\pi)^{-1} \in \{+1, -1\}.
\label{seq:inv_constraint_sc1}
\eg
If it is $+1$, both $\phi_{k_*} (0)$ and $\phi_{k_*} (\pi)$ are nonzero (up to accidental zeros), while if it is $-1$, both $\phi_{k_*} (0)$ and $\phi_{k_*} (\pi)$ are stable zeros.
Eq.~\eqref{seq:inv_constraint_sc1} implies $\xi_c (k_*) \xi_c (k_* + \pi)^{-1} = \xi_{\rm exc} (0) \xi_{\rm exc} (\pi)^{-1}$, which gives $x_c = x_{\rm exc} \pmod 1$ and hence $s_c = 0 \pmod 1$.

If instead exactly one of $\phi_{k_*} (0)$ or $\phi_{k_*} (\pi)$ is a stable zero, then the two constraints differ by a minus sign, so that
\bg
\xi_c (k_*) \, \xi_v (k_*)^{-1} \, \xi_{\rm exc} (0)^{-1}
= - \xi_c (k_* + \pi) \, \xi_v (k_*)^{-1} \, \xi_{\rm exc} (\pi)^{-1} \in \{+1, -1\}.
\label{seq:inv_constraint_sc2}
\eg
If it is $+1$, $\phi_{k_*} (0)$ is nonzero and $\phi_{k_*} (\pi)$ is a stable zero, while if it is $-1$, $\phi_{k_*} (0)$ is a stable zero and $\phi_{k_*} (\pi)$ is nonzero.
Eq.~\eqref{seq:inv_constraint_sc2} implies
\bg
\xi_c (k_*) \, \xi_c (k_* + \pi)^{-1} = e^{i \pi} \, \xi_{\rm exc} (0) \, \xi_{\rm exc} (\pi)^{-1},
\eg
which, using Eqs.~\eqref{seq:1d_berry_P} and \eqref{seq:wannier_def}, gives $x_c = x_{\rm exc} + 1/2 \pmod 1$ and hence $s_c = 1/2 \pmod 1$.

A completely analogous argument applies to the Wannier-center shift between valence band and exciton, $s_v = x_v - x_{\rm exc} \pmod 1$.
To see this, we rewrite Eq.~\eqref{seq:inv_constraint1} as
\bg
\left[ \xi_c (k_* + \pi) \, \xi_v (k_* + \pi)^{-1} \, \xi_{\rm exc} (0)^{-1} \right] \, \phi_{k_* + \pi} (0) = \phi_{k_* + \pi} (0),
\quad
\left[ \xi_c (k_* + \pi) \, \xi_v (k_*)^{-1} \, \xi_{\rm exc} (\pi)^{-1} \right] \, \phi_{k_*} (\pi) = \phi_{k_*} (\pi),
\label{seq:inv_constraint2}
\eg
If both $\phi_{k_* + \pi} (0)$ and $\phi_{k_*} (\pi)$ are stable zeros, or if both are nonzero, then
\bg
\xi_c (k_* + \pi) \, \xi_v (k_* + \pi)^{-1} \, \xi_{\rm exc} (0)^{-1}
= \xi_c (k_* + \pi) \, \xi_v (k_*)^{-1} \, \xi_{\rm exc} (\pi)^{-1} \in \{+1,-1\},
\eg
which implies $\xi_v (k_*)  \, \xi_v (k_* + \pi)^{-1}= \xi_{\rm exc} (0) \, \xi_{\rm exc} (\pi)^{-1}$.
This corresponds to $x_v = x_{\rm exc} \pmod 1$ and hence $s_v = 0 \pmod 1$.

If instead exactly one of $\phi_{k_* + \pi} (0)$ or $\phi_{k_*} (\pi)$ is a stable zero, then
\bg
\xi_c (k_* + \pi) \, \xi_v (k_* + \pi)^{-1} \, \xi_{\rm exc} (0)^{-1}
= - \xi_c (k_*+ \pi) \, \xi_v (k_*)^{-1} \, \xi_{\rm exc} (\pi)^{-1} \in \{+1,-1\},
\eg
which implies $\xi_v (k_*) \, \xi_v (k_* + \pi)^{-1} = e^{i \pi} \, \xi_{\rm exc} (0) \, \xi_{\rm exc} (\pi)^{-1}$.
Thus, $x_v = x_{\rm exc} + 1/2 \pmod 1$ and $s_v = 1/2 \pmod 1$.
Together, the results for $s_c$ and $s_v$ provide a complete classification of stable-zero patterns in 1D with inversion symmetry.

We now comment further on the structure and number of possible zero patterns.
Naively, one may expect $2^4=16$ possible zero patterns, since there are four EWF components at the HSM, $\phi_{k_*}(p_*)$ with $k_*,p_* \in \{0,\pi\}$, each of which can be either zero or nonzero.
However, not all such patterns are allowed.
The point is that we assume insulating bands and a gapped exciton, so that the Wannier centers, and hence the shifts $s_{c,v}$, are well defined.
This imposes consistency conditions on the zero patterns.
For example, consider a configuration of the form
\bg
\bbm
\phi_0(0) & \phi_\pi(0) \\
\phi_0(\pi) & \phi_\pi(\pi)
\ebm
= \bbm 0 & 0 \\
0 & \bullet \ebm,
\eg
where $\bullet$ denotes a nonzero value.
From the first column, one would infer $s_c = 0$, while from the second column one would infer $s_c = 1/2$, leading to a contradiction.
Thus, such patterns are excluded.
In general, the requirement that both the bands and the exciton are gapped, so that the associated topological invariants are well defined, restricts the allowed zero patterns.
A similar constraint will also appear in the 2D cases with rotation symmetry considered below.

It is useful to recast the above constraints in terms of a simple counting rule.
Let $N_{(k_*,p_*)} \in \{0,1\}$ be defined by $N_{(k_*,p_*)} = 1$ if $\phi_{k_*} (p_*)$ is a stable zero, and $N_{(k_*,p_*)} = 0$ otherwise.
Then the above analysis implies
\bg
2s_c = N_{(0,0)} + N_{(0,\pi)} = N_{(\pi,0)} + N_{(\pi,\pi)} \pmod 2,
\nn
2s_v = N_{(\pi,0)} + N_{(0,\pi)} = N_{(0,0)} + N_{(\pi,\pi)} \pmod 2.
\eg
In particular, this shows that the zero pattern is constrained beyond independent assignments of $N_{(k_*,p_*)}$.
Moreover, restricting to $p=0$, we only access $N_{(0,0)}$ and $N_{(\pi,0)}$, whose sum satisfies
\bg
N_{(0,0)} + N_{(\pi,0)} = 2s_c + 2s_v = 2(s_c - s_v) \pmod 2.
\eg
Using $s_c - s_v = x_c - x_v \pmod 1$, this determines the relative band Wannier center, $\dx := x_c - x_v \pmod 1$, directly from the $p=0$ zero pattern.

\subsection{$C_n$ rotation symmetry in two dimensions}
\label{app:zero_2d}
We now consider stable-zero patterns in 2D systems with $C_n$ rotation symmetry ($n = 2,3,4,6$).
The central quantities are the Chern numbers of the conduction and valence bands and the exciton, $\mc C_{c,v,\rm exc}$, from which we define the Chern-number shifts and the relative band Chern number,
\bg
\mc S_{c/v} = \mc C_{c/v} - \mc C_{\rm exc} \pmod n,
\quad
\dC = \mc C_c - \mc C_v = \mc S_c - \mc S_v \pmod n,
\label{seq:shift_def}
\eg
which capture the mismatch between the band and exciton Chern numbers, and the relative topology between the conduction and valence bands.
All topological invariants appearing below are understood modulo $n$, even when not explicitly indicated.

Before proceeding, we briefly summarize the main results.
In the $C_2$ case, each zero pattern uniquely determines $\mc C_{c,v,\rm exc}$, and hence also $(\mc S_c,\mc S_v)$ and $\dC$.
As a result, simple counting rules for the topological invariants, in terms of $N_{(\bk_*, \bp_*)}$ can be established (with $N_{(\bk_*, \bp_*)}=1$ if $\phi_{\bk_*} (\bp_*)$ is a stable zero, and $0$ otherwise).
In contrast, for $C_n$ with $n=3,4,6$, the correspondence between zero patterns and topology is weaker.
A given zero pattern can be compatible with multiple values of $\mc C_{c,v,\rm exc}$ and the derived quantities.
Nevertheless, as we will show, the zero pattern still encodes nontrivial and structured information, and certain patterns impose useful constraints on $\mc C_{c,v,\rm exc}$, $\dC$, and, in some cases, $(\mc S_c,\mc S_v)$.

We now outline the structure of the analysis.
We first focus on the $C_2$ case, where we derive explicit counting rules that determine the topological invariants directly from the zero pattern.
We then turn to the cases $C_n$ with $n = 3,4,6$, where we analyze the group-theoretical constraints underlying the stable-zero condition and, based on this structure, describe a systematic procedure to determine the relation between zero patterns and the topological invariants.
We further present representative examples that illustrate the range of possible constraints and their physical implications.
Throughout this subsection, the HSM are defined in Fig.~\ref{sfig:2D_BZ}, and we analyze the full zero pattern of the exciton EWF.
The case of $\bp = 0$ will be discussed separately in the next subsection.

\tocless{\subsubsection{$C_2$ symmetry: counting rules for topological invariants}}{}
We begin with the $C_2$ case.
As noted above, this case can be analyzed by extending the method used in 1D with inversion symmetry, together with the Thouless pump picture~\cite{thouless1983quantization}.
In particular, the Chern number modulo $2$ can be obtained from the change of the Wannier center (equivalently, polarization) along a 1D cut.
For example, taking the Wannier center along the $k_x$ direction and viewing $k_y$ as the pumping parameter, one may compare the Wannier centers at $k_y=0$ and $k_y=\pi$.
Equivalently, one may exchange the roles of $k_x$ and $k_y$, or consider other directions.
For instance, taking the polarization along the $k_x+k_y$ direction with $k_x-k_y$ as the pumping parameter leads to an analogous formulation, yielding the same result for gapped systems.
For notational simplicity, we use $(k_x, k_y)$ for both the bands and the exciton in this discussion, and distinguish $\bk$ and $\bp$ only when necessary.

In a 2D $C_2$-symmetric system, fixing $k_y=0$ or $\pi$ reduces the problem to an effective 1D subsystem with inversion symmetry, so that the 1D analysis discussed above can be directly applied.
More generally, the same applies to other $C_2$-symmetric cuts, such as those along the $k_x$ direction or along the $k_x \pm k_y$ directions.
Let us first focus on the cuts at $k_y=0$ and $\pi$.
At each of these values, one can define the Wannier centers for the conduction band, valence band, and exciton, $x_{c,v,\rm exc} (k_y) \pmod 1 \in \{0,1/2\}$, in the same way as in the 1D inversion-symmetric case.
Viewing the system as a 1D inversion-symmetric Thouless pump along $k_y$, the corresponding Chern numbers modulo $2$ are determined by the change of Wannier centers between $k_y=0$ and $\pi$, namely,
\bg
\mc C_a = 2 \left[ x_a (k_y=\pi)-x_a (k_y=0) \right] \pmod{2},
\eg
for $a=c,v, {\rm exc}$.
It is then useful to introduce their relative differences:
\bg
s_{c/v}^{(k_y=0)} = x_{c/v} (k_y=0) - x_{\rm exc} (k_y=0),
\quad
s_{c/v}^{(k_y=\pi)} = x_{c/v} (k_y=\pi) - x_{\rm exc} (k_y=\pi),
\nn
t_{c/v}^{(k_y=0)} = x_{c/v} (k_y=0) - x_{\rm exc} (k_y=\pi),
\quad
t_{c/v}^{(k_y=\pi)} = x_{c/v} (k_y=\pi) - x_{\rm exc} (k_y=0).
\eg
All these quantities are defined modulo $1$ and take values $0$ or $1/2$.
These quantities are related to the topological invariants via
\ba
\mc C_{\rm exc} =& 2 s_c^{(k_y=0)} + 2 t_c^{(k_y=0)}
= 2 s_v^{(k_y=0)} + 2 t_v^{(k_y=0)} \pmod 2,
\nn
\mc S_c =& 2 s_c^{(k_y=0)} + 2 s_c^{(k_y=\pi)}
= 2 t_c^{(k_y=0)} + 2 t_c^{(k_y=\pi)} \pmod 2,
\nn
\mc S_v =& 2 s_v^{(k_y=0)} + 2 s_v^{(k_y=\pi)}
= 2 t_v^{(k_y=0)} + 2 t_v^{(k_y=\pi)} \pmod 2,
\label{seq:thouless_ky}
\ea
and hence also to $\mc C_{c,v}$ and $\dC$.
Eq.~\eqref{seq:thouless_ky} both determines the topological invariants ($\mc C_{c,v,{\rm exc}}, \mc S_{c,v}, \dC$) and imposes constraints among $s_{c,v}^{(k_y)}$ and $t_{c,v}^{(k_y)}$.

Next, the quantities $s_{c,v}^{(k_y)}$ and $t_{c,v}^{(k_y)}$ can be related to the zero pattern through the symmetry constraints, following exactly the same logic as in the 1D case.
To this end, we evaluate Eq.~\eqref{seq:exc_sewing_phi} for $g=C_2$ at the HSM $\bk_*, \bp_* \in \{ \Gamma, X, Y, M \}$.
At these momenta, the sewing matrices reduce to $C_2$ eigenvalues, and the constraint becomes
\bg
\left[ B_{c, C_2} (\bk_*+\bp_*) \, B_{v, C_2}(\bk_*)^{-1} \, \mc B_{C_2} (\bp_*)^{-1} \right] \phi_{\bk_*} (\bp_*)
= \phi_{\bk_*} (\bp_*).
\eg
For example, taking $\bp_*=\Gamma, X$ and $\bk_*=\Gamma$, the resulting constraints have the same structure as those of the 1D inversion-symmetric case [Eq.~\eqref{seq:inv_constraint1}].
Therefore, one can directly apply the 1D analysis to relate $s_c^{(k_y=0)}$ to the zero pattern of $[\phi_{\Gamma} (\Gamma), \phi_{\Gamma} (X)]$.
In particular, $2 s_c^{(k_y=0)} = N_{(\Gamma,\Gamma)} + N_{(\Gamma,X)} \pmod 2$.
Repeating this analysis for all $\bk_*, \bp_*$, one obtains the full set of constraints associated with Thouless pumping along the $k_y$ direction:
\begin{equation}
\begin{alignedat}{2}
2 s_c^{(k_y=0)} =& N_{(\Gamma,\Gamma)} + N_{(\Gamma,X)}
= N_{(X, \Gamma)} + N_{(X,X)},
\quad
& 2 s_c^{(k_y=\pi)} =& N_{(\Gamma,Y)} + N_{(\Gamma,M)}
= N_{(X,Y)} + N_{(X,M)},
\\
2 t_c^{(k_y=0)} =& N_{(Y,Y)} + N_{(Y,M)}
= N_{(M,Y)} + N_{(M,M)},
\quad
& 2 t_c^{(k_y=\pi)} =& N_{(Y,\Gamma)} + N_{(Y,X)} = N_{(M,\Gamma)} + N_{(M,X)},
\\
2 s_v^{(k_y=0)} =& N_{(\Gamma,\Gamma)} + N_{(X,X)}
= N_{(\Gamma,X)} + N_{(X,\Gamma)},
\quad
& 2 s_v^{(k_y=\pi)} =& N_{(Y,Y)} + N_{(M,M)}
= N_{(Y,M)} + N_{(M,Y)},
\\
2 t_v^{(k_y=0)} =& N_{(\Gamma,Y)} + N_{(X,M)}
= N_{(X,Y)} + N_{(\Gamma,M)},
\quad
& 2 t_v^{(k_y=\pi)} =& N_{(Y,\Gamma)} + N_{(M,X)}
= N_{(Y,X)} + N_{(M,\Gamma)},
\end{alignedat}
\label{seq:thouless_ky_rule}
\end{equation}
where all equalities are understood modulo $2$.

If one considers only the constraints in Eq.~\eqref{seq:thouless_ky_rule}, certain zero patterns incompatible with a gapped system may still be allowed.
To eliminate such cases, one must also impose the analogous constraints associated with Thouless pumping along the $k_x$ direction, i.e., from the Wannier-center shifts at $k_x=0,\pi$, which can be derived in the same way.
The resulting relations are
\begin{equation}
\begin{alignedat}{2}
2 s_c^{(k_x=0)} =& N_{(\Gamma,\Gamma)} + N_{(\Gamma,Y)}
= N_{(Y,\Gamma)} + N_{(Y,Y)},
\quad
& 2 s_c^{(k_x=\pi)} =& N_{(\Gamma,X)} + N_{(\Gamma,M)}
= N_{(Y,X)} + N_{(Y,M)},
\\
2 t_c^{(k_x=0)} =& N_{(X,X)} + N_{(X,M)}
= N_{(M,X)} + N_{(M,M)},
\quad
& 2 t_c^{(k_x=\pi)} =& N_{(X,\Gamma)} + N_{(X,Y)}
= N_{(M,\Gamma)} + N_{(M,Y)},
\\
2 s_v^{(k_x=0)} =& N_{(\Gamma,\Gamma)} + N_{(Y,Y)}
= N_{(\Gamma,Y)} + N_{(Y,\Gamma)},
\quad
& 2 s_v^{(k_x=\pi)} =& N_{(X,X)} + N_{(M,M)}
= N_{(X,M)} + N_{(M,X)},
\\
2 t_v^{(k_x=0)} =& N_{(\Gamma,X)} + N_{(Y,M)}
= N_{(\Gamma,M)} + N_{(Y,X)},
\quad
& 2 t_v^{(k_x=\pi)} =& N_{(X,\Gamma)} + N_{(M,Y)}
= N_{(X,Y)} + N_{(M,\Gamma)},
\end{alignedat}
\label{seq:thouless_kx_rule}
\end{equation}
which hold modulo $2$.

Finally, for a gapped system, the topological invariants $[\mc S_c, \mc S_v, \mc C_{\rm exc} \pmod 2]$ can be extracted directly from the zero pattern.
For example,
\ba
\mc S_c =& 2 s_c^{(k_y=0)} + 2 s_c^{(k_y=\pi)}
= \left[ N_{(\Gamma,\Gamma)} + N_{(\Gamma,X)} \right] + \left[ N_{(\Gamma,Y)} + N_{(\Gamma,M)} \right] \pmod 2,
\nn
\mc S_v =& 2 s_v^{(k_y=0)} + 2 s_v^{(k_y=\pi)}
= \left[ N_{(\Gamma,\Gamma)} + N_{(X,X)} \right]
+ \left[ N_{(Y,Y)} + N_{(M,M)} \right] \pmod 2,
\nn
\mc C_{\rm exc} =& 2 s_c^{(k_y=0)} + 2 t_c^{(k_y=0)}
= \left[ N_{(\Gamma,\Gamma)} + N_{(\Gamma,X)} \right] + \left[ N_{(Y,Y)} + N_{(Y,M)} \right] \pmod 2.
\ea
Therefore, it follows that $\mc C_{c,v} \pmod 2$ can also be directly determined from the zero pattern.
Moreover, the relative band Chern number $\dC$ modulo 2 can already be inferred from the $\bp_*=0$ zero pattern alone:
\bg
\dC = \mc C_c - \mc C_v
= N_{(\Gamma,\Gamma)} + N_{(X,\Gamma)} + N_{(Y,\Gamma)} + N_{(M,\Gamma)} \pmod 2.
\label{seq:c2_rel_Ch_rule}
\eg
To see this, note first that
\bg
N_{(\Gamma,\Gamma)} + N_{(X,\Gamma)} + N_{(Y,\Gamma)} + N_{(M,\Gamma)} + \mc S_c + \mc S_v
= \square_{(\Gamma,X)} + \square_{(\Gamma,Y)} + \square_{(\Gamma, M)}
\eg
can be expressed as the sum of three quantities $\square_{(\Gamma,\bk_*)}$ associated with $(\Gamma,X)$, $(\Gamma,Y)$, and $(\Gamma,M)$:
\bg
\square_{(\Gamma,\bk_*)} = N_{(\Gamma,\Gamma)} + N_{(\Gamma,\bk_*)} + N_{(\bk_*,\Gamma)} + N_{(\bk_*,\bk_*)} \pmod 2
\eg
for $\bk_*=X,Y,M$.
The first two, $\square_{(\Gamma,X)}$ and $\square_{(\Gamma,Y)}$, vanish due to the constraints [Eqs.~\eqref{seq:thouless_ky_rule} and \eqref{seq:thouless_kx_rule}] obtained from the Thouless pumps of the $k_y$- and $k_x$-directed Wannier centers.
The last one, $\square_{(\Gamma,M)}$, vanishes analogously by considering the $[11]$-directed Thouless pump, i.e., the 1D subsystem oriented along $k_x+k_y$ with pumping parameter $k_x-k_y$.
In particular, the zero patterns of $[\phi_\Gamma (\Gamma), \phi_\Gamma (M)]$ and $[\phi_M (\Gamma), \phi_M (M)]$ are related to the corresponding $[11]$-directed Wannier-center shift, from which it follows that $\square_{(\Gamma,M)} = 0 \pmod 2$.
This establishes Eq.~\eqref{seq:c2_rel_Ch_rule}.
More generally, analogous results that extract topological information from the $\bp_*=0$ zero pattern alone for other $C_n$ symmetries are discussed in Sec.~\ref{app:zero_p0}.
\\

\tocless{\subsubsection{$C_{3,4,6}$ symmetries: structure of stable zeros and inference of topology}}{}
Unlike the $C_2$ case, where the full zero pattern admits simple counting rules that completely determine the relevant topological invariants (namely $\mc C_{c,v,\rm exc}$, and hence $\mc S_{c,v}$ and $\dC$), higher rotation symmetries $C_{n=3,4,6}$ do not admit such simple rules.
In general, a given zero pattern is compatible with multiple values of these invariants, and therefore does not uniquely determine them.
Nevertheless, the zero pattern remains a useful diagnostic.
In particular, there exist patterns that constrain the relative band topology $\dC$ and the exciton Chern number $\mc C_{\rm exc}$ in a meaningful way, even if a unique determination is not always possible.
By contrast, the Chern-number shifts $\mc S_{c,v}$ are in general more weakly constrained.
Rather than listing an exhaustive classification of zero patterns, which can be obtained by an algorithmic enumeration, we instead discuss the group-theoretic properties of the stable-zero condition and how these can be used to guide such an enumeration.
Below, we first discuss the group-theoretic structure of the stable-zero condition.
We then describe a procedure to obtain the mapping from zero patterns to topological invariants.
Finally, we discuss the predictive power of zero patterns and present representative examples that illustrate the underlying structures.

\paragraph{Group-theoretic structure of the stable-zero condition:}
We begin by restating the symmetry constraint [Eq.~\eqref{seq:exc_sewing_phi}] on the EWF.
For a rotation $g \in C_n$, the EWF satisfies
\bg
\phi_{g \bk}(g \bp) = \Lambda_g (\bk, \bp)\,\phi_ \bk (\bp),
\quad
\Lambda_g (\bk, \bp) = B_{c, g} (\bk + \bp) \, B_{v, g} (\bk)^{-1} \, \mc B_g (\bp)^{-1}.
\label{seq:zero_cond_group}
\eg
The coefficient $\Lambda_g (\bk, \bp)$ inherits the group multiplication law
\bg
\Lambda_{g_1 g_2} (\bk, \bp) = \Lambda_{g_1} (g_2 \bk,  g_2 \bp) \, \Lambda_{g_2} (\bk, \bp),
\eg
from that of the sewing matrices [Eq.~\eqref{seq:sewing_rule}].
This structure leads to two general consequences.

First, suppose that we consider HSM $\bk_*$ and $\bp_*$ such that $g \bk_* = \bk_*$ and $g \bp_* = \bp_*$, where $g$ is chosen to be a generator of the common little group of $\bk_*$ and $\bp_*$.
(In the cases considered here, the little group is cyclic and generated by a single element.
For certain choices of $\bk_*$ and $\bp_*$, this generator can be $g = {\rm id}$, the identity operation.)
Then $\bk$, $\bp$, and $\bk + \bp$ are all fixed points of $g$, and the symmetry constraint takes the local form:
\bg
\phi_{\bk_*} (\bp_*) = \Lambda_g (\bk_*, \bp_*) \, \phi_{\bk_*} (\bp_*).
\label{seq:zero_cond}
\eg
In this case, it is sufficient to consider this single generator $g$ to determine the stable-zero condition at $(\bk_*, \bp_*)$.
Constraints from $g^{-1}$ or higher powers of $g$ do not provide additional information, as they follow from repeated application of the above relation.
For example, when $g \bk_*=\bk_*$ and $g\bp_*=\bp_*$, one has $\Lambda_{g^{-1}} (\bk_*, \bp_*) = \Lambda_g (\bk_*, \bp_*)^{-1}$, so the constraint from $g^{-1}$ follows directly from that of $g$.
More generally, constraints from $g^m$ follow from repeated application of the same relation.

Second, we consider the orbit generated by the $C_n$ rotation.
Starting from $(\bk_*,\bp_*)$, repeated application of $g = C_n$ generates the orbit of $(\bk_*,\bp_*)$ under $C_n$, relating $\phi_{\bk_*} (\bp_*)$ to amplitudes at other points in the orbit.
Since the orbit is finite, one eventually returns to $(\bk_*,\bp_*)$, yielding
\bg
\phi_{\bk_*} (\bp_*) = \Lambda_{g^\ell} (\bk_*,\bp_*) \, \phi_{\bk_*} (\bp_*),
\eg
where $g^\ell$ is the smallest power of $g$ that fixes both $\bk_*$ and $\bp_*$.
This element necessarily belongs to the common little group of $\bk_*$ and $\bp_*$.
If $\Lambda_{g^\ell} (\bk_*,\bp_*) \neq 1$, the orbit closure produces a nontrivial local constraint on $\phi_{\bk_*} (\bp_*)$.
If $\Lambda_{g^\ell} (\bk_*,\bp_*) = 1$, the resulting constraint is trivial and does not impose any restriction on $\phi_{\bk_*} (\bp_*)$.
This always occurs when the common little group of $\bk_*$ and $\bp_*$ is trivial, in which case $g^\ell = (C_n)^n = {\rm id}$.

\paragraph{Algorithmic determination of zero-pattern constraints:}
The basic idea is straightforward.
For each allowed assignment of symmetry eigenvalues of the conduction band, valence band, and exciton at the HSM, one can determine both the corresponding topological invariants and the associated zero pattern.
More explicitly, from a given set of symmetry eigenvalues, Eq.~\eqref{seq:chern_formula} yields $\mc C_{c,v,\rm exc}$ and thus the derived quantities $\mc S_{c,v}$ and $\dC$.
The same symmetry eigenvalues also determine, through the stable-zero condition [Eq.~\eqref{seq:zero_cond}], which entries $\phi_{\bk_*} (\bp_*)$ must vanish.
Repeating this procedure for all possible sets of symmetry eigenvalues yields a complete dataset.
This produces a map from the zero pattern to the set of compatible topological invariants.

The group-theoretic structure discussed above makes this enumeration considerably more efficient.
For $C_3$, all HSM share the same little group generated by $C_3$.
Therefore, for each entry $\phi_{\bk_*} (\bp_*)$, it is sufficient to check the stable-zero condition from $g=C_3$ alone, since the condition from $C_3^2=C_3^{-1}$ is redundant.
Thus, for each symmetry eigenvalue configuration, the full zero pattern is obtained by evaluating the $C_3$ constraint at all $(\bk_*, \bp_*) \in \{\Gamma, K, K'\} \times \{\Gamma, K, K'\}$.

For $C_{n=4,6}$, by contrast, the HSM do not all share the same little group, and one must additionally account for the orbit structure under $C_n$.
It is convenient to organize the enumeration by $C_n$ orbits of $(\bk_*,\bp_*)$, since entries in the same orbit are related by symmetry and therefore need not be checked independently.
Moreover, once an orbit closes, the resulting relation reduces to a local constraint associated with the common little group, as discussed above.
In this way, the zero pattern can be determined orbit by orbit, rather than entry by entry.

For concreteness, we illustrate this procedure for $C_6$.
For $(\bk_*,\bp_*) \in \{\Gamma\} \times \{\Gamma\}$, i.e., $\bk_*=\bp_*=\Gamma$, one only needs the $C_6$ condition.
For $(\bk_*,\bp_*) \in \{\Gamma, K, K'\} \times \{\Gamma, K, K'\}$, the relevant local condition is the $C_3$ condition, since $C_3$ generates the common little group of these points.
In particular, for $(\bk_*,\bp_*) = (\Gamma,\Gamma)$, the $C_3$ condition does not provide any additional constraint beyond the $C_6$ condition.

Within this sector, symmetry further relates entries in the same orbit.
For example,
\bg
(\Gamma,K) \xrightarrow{C_6} (\Gamma,K') \xrightarrow{C_6} (\Gamma,K),
\quad
(K,\Gamma) \xrightarrow{C_6} (K',\Gamma) \xrightarrow{C_6} (K,\Gamma),
\nn
(K,K) \xrightarrow{C_6} (K',K') \xrightarrow{C_6} (K,K),
\quad
(K,K') \xrightarrow{C_6} (K',K) \xrightarrow{C_6} (K,K'),
\eg
so that, for instance, if $\phi_{\Gamma} (K)$ is a stable zero, then $\phi_{\Gamma} (K')$ is also a stable zero.
Similarly, for $(\bk_*, \bp_*) \in \{\Gamma, M, M', M''\} \times \{\Gamma, M, M', M''\}$, one organizes the entries into length-three $C_6$ orbits, while the associated local constraint is the $C_2$ condition.
Finally, when $K$-type and $M$-type momenta are mixed, one obtains length-six $C_6$ orbits.
In this case the common little group is trivial, so the orbit closure yields only the identity constraint and no nontrivial stable-zero condition.

\paragraph{Topological information encoded in zero patterns and examples:}
We now summarize what topological information can be inferred from the full zero pattern in $C_{3,4,6}$-symmetric systems.
We first illustrate the structure through explicit examples, and then summarize the general constraints.
We begin with explicit examples in the $C_4$ case.
Due to symmetry, it is sufficient to consider representatives of each orbit of $(\bk_*,\bp_*)$.
Accordingly, the zero pattern can be specified by the following ten entries:
\ba
\left[ \phi_{\Gamma} (\Gamma), \, \phi_M (\Gamma),
\, \phi_{\Gamma} (M), \, \phi_M (M);
\, \phi_X (\Gamma), \, \phi_X (M),
\, \phi_{\Gamma} (X), \, \phi_M (X);
\, \phi_X (X), \, \phi_X (Y) \right].
\ea

(i) We first consider an example that illustrates the non-uniqueness of the mapping from zero patterns to topological invariants.
Consider the pattern
\ba
\left[ 0, \, 0, \, 0, \, 0; \, 0, \, 0, \, 0, \, 0; \, \bullet, \, \bullet \right],
\ea
where $0$ and $\bullet$ denote stable zeros and generically nonvanishing entries, respectively.
This pattern can arise from different set of symmetry eigenvalues.
For example, for one choice of symmetry eigenvalues at $(\Gamma, M, X)$ for the conduction/valence bands and the exciton:
\ba
c: \, (1,1,-1), \quad v: \, (1,1,1), \quad {\rm exc}: \, (-1,-1,+1),
\ea
where in each triplet the first two entries correspond to $C_4$ eigenvalues at $\Gamma$ and $M$, and the last to the $C_2$ eigenvalue at $X$.
Then, from Eq.~\eqref{seq:chern_formula}, one finds $(\mc C_c, \mc C_v, \mc C_{\rm exc}) = (2,0,0)$ modulo $4$, and hence $(\mc S_c, \mc S_v) = (2,0)$ and $\dC = 2$ modulo $4$.
On the other hand, for a different assignment of symmetry eigenvalues,
\ba
c: \, (1,1,1), \quad v: \, (1,1,-1), \quad {\rm exc}: \, (-1,-1,-1),
\ea
one instead obtains $(\mc C_c, \mc C_v, \mc C_{\rm exc}) = (0, 2, 2)$ modulo $4$, leading to $(\mc S_c, \mc S_v) = (2,0)$ and $\dC = 2$ modulo $4$.
Thus, the same zero pattern can correspond to distinct values of $\mc C_{c,v,\rm exc}$, and therefore does not uniquely determine these invariants.
At the same time, the relative band Chern number $\dC$ and the shifts $(\mc S_c, \mc S_v)$ are fixed to $\dC = 2$ and $(\mc S_c, \mc S_v)=(2,0)$ modulo $4$, illustrating that the zero pattern can still impose nontrivial constraints on certain combinations of invariants.

(ii) We next present examples where the zero pattern imposes nontrivial constraints on $\mc C_{\rm exc}$.
In the $C_4$ case, there exist patterns that restrict $\mc C_{\rm exc}$ to the subsets $\{0,2\}$ or $\{1,3\}$ modulo $4$.
In other words, the zero pattern determines a subset of allowed values of $\mc C_{\rm exc}$, but does not uniquely fix its value within that subset.
For instance, a pattern that implies $\mc C_{\rm exc} \in \{1,3\}$ does not distinguish between $\mc C_{\rm exc}=1$ and $\mc C_{\rm exc}=3$, but one can infer that $\mc C_{\rm exc} \notin \{0,2\}$.
Among the many possible zero patterns, the following representative examples constrain $\mc C_{\rm exc} \in \{0,2\}$:
\bg
\dots, \, \left[0, \, 0, \, 0, \, 0; \, 0, \, 0, \, 0, \, 0; \, 0, \, 0 \right], \,
\left[0, \, 0, \, 0, \, 0; \, 0, \, 0, \, 0, \, \bullet; \, 0, \, 0 \right],
\, \dots,
\eg
Similarly, the following patterns constrain $\mc C_{\rm exc} \in \{1,3\}$:
\bg
\dots, \, \left[0, \, 0, \, 0, \, 0; \, 0, \, \bullet, \, 0, \, 0; \, 0, \, 0 \right], \,
\left[0, \, 0, \, 0, \, \bullet; \, 0, \, \bullet, \, 0, \, 0; \, 0, \, 0 \right],
\, \dots.
\eg
These examples demonstrate that, although $\mc C_{\rm exc}$ is not uniquely determined in general, the zero pattern can still reduce the set of allowed values.

(iii) Finally, we consider examples where the zero pattern constrains the Chern number shifts $(\mc S_c, \mc S_v)$.
Since the underlying Chern numbers $(\mc C_c, \mc C_v, \mc C_{\rm exc})$ are not uniquely determined in general, the resulting constraints on $(\mc S_c, \mc S_v)$ are less restrictive.
Nevertheless, certain zero patterns still enforce nontrivial relations between $\mc S_c$ and $\mc S_v$.
A particularly interesting case is when the pattern enforces $\mc S_c = \mc S_v \neq 0$.
If one additionally knows that $\mc C_c = \mc C_v = 0$, this immediately implies $\mc C_{\rm exc} = -\mc S_c = - \mc S_v \neq 0$, indicating a topological exciton formed on top of topologically trivial bands.
In the $C_4$ case, such patterns can, for example, enforce $(\mc S_c, \mc S_v) = (2,2)$ or $(\mc S_c,\mc S_v) \in \{(1,1), (3,3)\}$.
Representative patterns realizing these cases include
\bg
(\mc S_c, \mc S_v) \in \{ (2,2) \} : \quad
\dots, \, \left[0, \, 0, \, 0, \, 0; \, \bullet, \, \bullet, \, 0, \, 0; \, 0, \, 0 \right], \,
\left[\bullet, \, \bullet, \, \bullet, \, \bullet; \, \bullet, \, \bullet, \, 0, \, 0; \, 0, \, 0 \right],
\, \dots,
\nn
(\mc S_c, \mc S_v) \in \{ (1,1), (3,3) \} : \quad
\dots, \, \left[0, \, 0, \, 0, \, 0; \, 0, \, \bullet, \, 0, \, 0; \, 0, \, 0 \right], \,
\left[0, \, 0, \, \bullet, \, \bullet; \, 0, \, \bullet, \, 0, \, 0; \, 0, \, 0 \right],
\, \dots.
\eg
These examples illustrate that certain zero patterns can still encode physically meaningful information about the Chern number shifts $\mc S_{c,v}$.

\paragraph{Summary:}
We conclude by summarizing the possible constraints on the mod-$n$ Chern numbers inferred from the full zero pattern in $C_n$-symmetric systems.
In the $C_3$ case, for a given pattern, the allowed values of $\mc C_{c,v,\rm exc}$ and $\dC$ are restricted to one of the following subsets:
\bg
\{0\}, \, \{1,2\}, \, \{0,1,2\}.
\eg
Here and below, the notation $\mc C_{c,v,\rm exc}$ indicates that each of $\mc C_c$, $\mc C_v$, and $\mc C_{\rm exc}$ independently takes values in one of the listed subsets.
In the $C_4$ case, for a given pattern, one finds
\bg
\mc C_{c,v,\rm exc} \in \{0,2\}, \, \{1,3\},
\quad
\dC \in \{0\}, \, \{2\}, \, \{0,2\}, \, \{1,3\}.
\eg
In the $C_6$ case, for a given pattern, both $\mc C_{c,v,\rm exc}$ and $\dC$ are restricted to one of the subsets
\bg
\{0\}, \, \{3\}, \, \{1,5\}, \, \{2,4\}, \, \{0,2,4\}, \, \{1,3,5\}.
\eg
While the constraints on $(\mc S_c,\mc S_v)$ are in general more intricate, as illustrated in the examples above, the zero pattern can still provide physically meaningful information on these quantities in specific cases.

\subsection{Complete list of $p = 0$ zero patterns under $C_n$}
\label{app:zero_p0}
Here, we list all possible $\bp=0$ (i.e., $\bp=\Gamma$) zero patterns and the corresponding constraints on the relative band Chern number $\dC$ in 2D $C_n$-symmetric systems.
Throughout this subsection, $\dC$ is understood modulo $n$.

(i) In $C_2$ symmetric systems, the $p=0$ zero pattern is specified by $[ \phi_{\Gamma} (\Gamma), \phi_X (\Gamma), \phi_Y (\Gamma), \phi_M (\Gamma)]$.
The possible patterns and the inferred $\dC$ are:
\ba
\dC = 0 :& \,\, [0,0,0,0], \, [0,0,\bullet,\bullet], \, [0,\bullet,0,\bullet], \, [0,\bullet,\bullet,0], \, [\bullet,0,0,\bullet], \, [\bullet,0,\bullet,0], \, [\bullet,\bullet,0,0], \, [\bullet,\bullet,\bullet,\bullet], \nn
\dC = 1 :& \,\, [0,0,0,\bullet], \, [0,0,\bullet,0], \, [0,\bullet,0,0], \, [0,\bullet,\bullet,\bullet], \, [\bullet,0,0,0], \, [\bullet,0,\bullet,\bullet], \, [\bullet,\bullet,0,\bullet], \, [\bullet,\bullet,\bullet,0].
\ea
Here, $0$ denotes a stable zero, and $\bullet$ a generically nonvanishing value.

(ii) In $C_3$ symmetric systems, the $p=0$ pattern is given by $[ \phi_{\Gamma} (\Gamma), \phi_K (\Gamma), \phi_{K'} (\Gamma)]$.
The corresponding constraints on $\dC$ are:
\ba
& \dC = 0 : \,\, [\bullet,\bullet,\bullet], \quad
\dC \in \{1,2\} : \,\, [0,\bullet,\bullet], \, [\bullet,0,\bullet], \, [\bullet,\bullet,0],
\nn
&\dC \in \{0,1,2\} : \,\, [0,0,0], \, [\bullet,0,0], \, [0,\bullet,0], \, [0,0,\bullet].
\ea
The last type provides no useful constraint on $\dC$.

(iii) In $C_4$ symmetric systems, the $p=0$ pattern can be written as $[\phi_{\Gamma} (\Gamma), \phi_M (\Gamma), \phi_X (\Gamma)]$ since $\phi_Y (\Gamma)$ is related to $\phi_X (\Gamma)$ by $C_4$ symmetry.
The corresponding values of $\dC$ inferred from these patterns are classified as follows:
\ba
& \dC = 0 : \,\, [\bullet,\bullet,\bullet], \quad
\dC = 2 : \,\, [\bullet,\bullet,0], \quad
\dC \in \{1,2,3\} : \,\, [0,\bullet,\bullet], \, [\bullet,0,\bullet],
\nn
& \dC \in \{0,1,3\} : \,\, [0,\bullet,0], \, [\bullet,0,0], \quad
\dC \in \{0,1,2,3\} : \,\, [0,0,0], \, [0,0,\bullet].
\ea

(iv) In $C_6$ symmetric systems, the $p=0$ pattern is given by $[\phi_{\Gamma} (\Gamma), \phi_K (\Gamma), \phi_M (\Gamma)]$.
The possible patterns and the inferred $\dC$ are:
\ba
& \dC = 0 : \,\, [\bullet,\bullet,\bullet], \quad
\dC = 3 : \,\, [\bullet,\bullet,0], \quad
\dC \in \{1,5\} : \,\, [\bullet,0,0],
\nn
& \dC \in \{2,4\} : \,\, [\bullet,0,\bullet], \quad
\dC \in \{1,2,3,4,5\} : \,\, [0,\bullet,\bullet],
\nn
& \dC \in \{0,1,2,4,5\} : \,\, [0,\bullet,0], \quad
\dC \in \{0,1,2,3,4,5\} : \,\, [0,0,0],\ [0,0,\bullet].
\ea

\section{Lattice model and numerical details}
\label{app:model}
In this section, we present the 1D lattice model used in the main text and explain how the exciton spectrum and EWF amplitude shown in Fig.~1 of the main text are obtained.
We also summarize the projected-Hamiltonian formula used in the numerics.
The general notation for tight-binding Hamiltonians, band eigenstates, and sewing matrices has already been introduced in Sec.~\ref{app:notation}.

\paragraph{Noninteracting lattice model:}
We consider a 1D lattice model with two sublattices (or orbitals) per unit cell, labeled by $i=1,2$.
For convenience, we take the intracell positions to satisfy $x_1=x_2=0$.
We start from the noninteracting Hamiltonian
\ba
\hat H_0 =& t_1 \sum_R \, (c^\dg_{R,1} c_{R,2} + c^\dg_{R,2} c_{R,1}) + t_2 \sum_R \, (c^\dg_{R+1,1} c_{R,2} + c^\dg_{R,2} c_{R+1,1})
\nn
&+ t_3 \sum_R \, (c^\dg_{R+2,1} c_{R,2} + c^\dg_{R,1} c_{R+2,2} + c^\dg_{R+2,2} c_{R,1} + c^\dg_{R,2} c_{R+2,1}).
\ea
This model was studied in Ref.~\cite{davenport2026exciton} in the context of topological excitons in 1D with inversion and spacetime inversion symmetries.
After Fourier transformation, the momentum-space Hamiltonian $H(k)$ is
\bg
\hat H_0 = \sum_k \sum_{i,j=1,2} \, H(k)_{ij} \, c^\dg_{k,i} c_{k,j},
\nn
H(k) = \bpm
0 & t_1 + t_2 Q_x + t_3 \left(Q_x^2 + Q_x^{-2}\right) \\
t_1 + t_2 Q_x^{-1} + t_3 \left(Q_x^2 + Q_x^{-2}\right) & 0 \epm,
\eg
where $Q_x = e^{-ik}$.
The two noninteracting bands are denoted by $v$ and $c$, where $v$ is the lower band and $c$ is the upper band.
Their energies and eigenstates are denoted by $\vep_{k,n}$ and $(u_{k,n})_i$, with $n=c,v$.
We use the parameter choice $(t_1,t_2,t_3) = (0.6,\,0.03,\,0.01)$.
The longer-range hopping $t_3$ is introduced so that, after adding the interaction below, the lowest exciton remains isolated from the other exciton branches for all total exciton momentum $p$.
In other words, it is used to obtain a gapped lowest exciton band.

The model has inversion symmetry represented by $U_P = \sg_x$.
At the inversion-invariant momenta $k=0,\pi$, the inversion eigenvalues are $\xi_v(0)=-1$, $\xi_v(\pi)=-1$, $\xi_c(0)=+1$, $\xi_c(\pi)=+1$.
Therefore, by Eq.~\eqref{seq:1d_berry_P}, both bands have zero Berry phase, or equivalently zero Wannier center: $x_v=x_c=0 \pmod 1$.

\paragraph{Interaction term:}
We next add a density-density interaction
\bg
\hat V = u_1 \sum_R \, n_{R,1} n_{R,2} + u_2 \sum_R \, n_{R+1,1} n_{R,2},
\eg
where $n_{R,i}=c^\dg_{R,i} c_{R,i}$.
We use $(u_1,u_2) = (0.1, 0.15)$.
Equivalently, writing the interaction explicitly in a symmetric form,
\bg
\hat V  = \frac{u_1}{2} \sum_R \, \left(
n_{R,1} n_{R,2} + n_{R,2} n_{R,1} \right)
+ \frac{u_2}{2} \sum_R \, \left( n_{R+1,1} n_{R,2} + n_{R,2} n_{R+1,1} \right).
\eg
This may be written more generally as
\bg
\hat V = \sum_{R,R',i,j} \, U_{R-R',i,j} \, n_{R,i} n_{R',j},
\eg
with the Fourier transform
\bg
\hat V = \sum_{p,q,q'} \sum_{i,j} \, U_{p,i,j} \, c^\dg_{p+q,i} c_{q,i} c^\dg_{q',j} c_{p+q',j},
\quad
U_{p,i,j} = \frac{1}{N_{\rm cell}} \sum_R \, U_{R,i,j} e^{-ipR}.
\eg

\paragraph{Projected Hamiltonian for the exciton:}
As explained in Sec.~\ref{app:ewf}, we study the exciton using the projected-Hamiltonian formalism.
The exciton basis is $\ket{k,p} = c^\dg_{k+p,c} c_{k,v} \ket{\gs}$, where $c^\dg_{k,c}$ and $c^\dg_{k,v}$ are the creation operators for the conduction and valence bands, respectively.
Then, the projected Hamiltonian is defined by
\bg
\mc H_{k,k'} (p) = \bra{k,p} \hat H \ket{k',p},
\quad
\hat H = \hat H_0 + \hat V.
\eg
The EWF $\phi_k^{(a)} (p)$ and exciton energy $E_{{\rm exc},a} (p)$ are obtained from
\bg
\sum_{k'} \, \mc H_{k,k'} (p) \, \phi_{k'}^{(a)} (p)
= E_{{\rm exc},a} (p) \, \phi_k^{(a)} (p).
\eg
Here $a$ is the exciton-band index.
In the main text we focus on the lowest exciton branch, so below the index $a$ may be omitted.

Following Ref.~\cite{davenport2026exciton}, one can express the projected interaction matrix elements in terms of the band-basis interaction tensor.
For convenience, we summarize only the final formulas here and refer to Ref.~\cite{davenport2026exciton} for the derivation.
Defining
\bg
U^{p,q,q'}_{n,n'm,m'} = \sum_{i,j} \, U_{p,i,j} \, \cm{(u_{p+q,n})_i} \, (u_{q,n'})_i \, \cm{(u_{q',m})_j} \, (u_{p+q',m'})_j,
\eg
where $n,n',m,m' \in \{c,v\}$ are band indices, the interaction Hamiltonian can be written in the band basis as
\bg
\hat V = \sum_{p,q,q'} \sum_{n,n',m,m'} \, U^{p,q,q'}_{n,n',m,m'} \, c^\dg_{p+q,n} c_{q,n'}
c^\dg_{q',m} c_{p+q',m'}.
\eg
In the exciton basis $\ket{k,p}$, the projected matrix element $\bra{k,p} \hat V \ket{k',p}$ reduces to
\bg
U^{p,k,k'}_{c,v,v,c} + U^{-p,k'+p,k+p}_{v,c,c,v} - U^{k'-k,k,k+p}_{v,v,c,c} - U^{k-k',k'+p,k'}_{c,c,v,v}
+ \delta_{k,k'} \sum_q \, \big( U^{q,k,k}_{v,v,v,v} - U^{0,k,q}_{v,v,v,v} - U^{0,q,k}_{v,v,v,v}
\nn
+ U^{k+p-q,q,q}_{c,c,c,c} + U^{0,k+p,q}_{c,c,v,v} + U^{0,q,k+p}_{v,v,c,c} - U^{q,k-q,k-q}_{v,c,c,v} - U^{q,k+p,k+p}_{v,c,c,v} \big)
+ \delta_{k,k'} \sum_{q,q'} \, \left( U^{0,q,q'}_{v,v,v,v} + U^{q,q',q'}_{v,c,c,v} \right).
\label{seq:app_projV}
\eg
The noninteracting contribution is
\bg
\bra{k,p} \hat H_0 \ket{k',p} = \delta_{k,k'} \, (\vep_{p+k,c} - \vep_{k,v} + \vep_{\gs}),
\label{eq:app_projH0}
\eg
where $\vep_{\gs} = \sum_k \, \vep_{k,v}$.

Finally, the last term in Eq.~\eqref{seq:app_projV} is equal to the interaction-induced shift of the ground-state energy,
\bg
\bra{\gs} \hat V \ket{\gs} = \sum_{q,q'} \, \left( U^{0,q,q'}_{v,v,v,v} + U^{q,q',q'}_{v,c,c,v} \right).
\label{seq:app_gs_shift}
\eg
Since we are interested in the excitation energy above the interacting ground state, this contribution, $\bra{\gs} \hat H \ket{\gs} = \bra{\gs} \hat V \ket{\gs} + \vep_\gs$, is subtracted from the projected exciton Hamiltonian.
Therefore, the spectrum shown in Fig.~1 of the main text is obtained from
\bg
\td{\mc H} (p)_{k,k'} = \bra{k,p} \hat H \ket{k',p} - \delta_{k,k'} \bra{\gs} \hat H \ket{\gs}.
\eg

\paragraph{Numerical results:}
For each total exciton momentum $p$, we diagonalize the matrix $\mc H (p)$ and obtain the exciton energies $E_{{\rm exc},a} (p)$ and EWFs $\phi_k^{(a)} (p)$.
The exciton spectrum plotted in Fig.~1 of the main text is the lowest branch $E_{\rm exc} (p)$ from $\td{\mc H} (p)$.
The quantity $|\phi_k (p)|^2$ shown in Fig.~1 of the main text is obtained from the normalized eigenvector of this lowest branch.

We are particularly interested in the topology of the lowest exciton band.
Using Eq.~\eqref{seq:exc_sewing_hsm}, its inversion eigenvalues at $p=0,\pi$ are found to be $\xi_{\rm exc} (0)=-1$ and $\xi_{\rm exc} (\pi)=+1$.
Therefore, the exciton Berry phase is $\pi$, or equivalently the exciton Wannier center is $x_{\rm exc} = \frac12 \pmod 1$.
Since the conduction and valence bands satisfy $x_c = x_v = 0 \pmod 1$, the Wannier-center shifts are $s_c = s_v = \frac12 \pmod 1$.
As discussed in Sec.~\ref{app:zero_1d}, the numerically obtained stable-zero pattern
\bg
\phi_0 (\pi) = \phi_\pi (\pi)=0,
\eg
is consistent with the topological data, namely the inversion eigenvalues of the conduction and valence bands and the exciton, and corresponds to one of two allowed patterns for $(s_c,s_v) = \left( \frac12, \frac12 \right)\pmod 1$, as classified in Sec.~\ref{app:zero_1d}.

\end{document}